\documentclass[11pt,a4paper]{article}

\usepackage{jcappub}
\usepackage[margin=0.95in]{geometry}

\usepackage[utf8]{inputenc}
\usepackage{graphicx}
\usepackage{verbatim}
\usepackage{epsfig}
\usepackage{subfigure}
\usepackage{color}
\usepackage{multirow}
\usepackage{comment}
\usepackage{slashed}
\usepackage{amsmath}
\usepackage{ulem}
\usepackage{framed}

\usepackage{amssymb,amsmath,amsfonts,mathrsfs}
\usepackage[dvipsnames]{xcolor}
\usepackage{graphicx}
\usepackage{longtable}
\usepackage{verbatim}
\usepackage{color}
\usepackage[mathscr]{euscript}
\usepackage{framed}
\usepackage{mdframed}  

\usepackage{cancel}
\usepackage{color}
\usepackage{hyperref}
\hypersetup{colorlinks, citecolor=bluscuro, linkcolor=black, urlcolor=bluscuro}
\definecolor{rossos}{cmyk}{0,1,1,0.55}
\definecolor{bluscuro}{rgb}{0.15, 0.2, .85}
\definecolor{bluchiaro}{cmyk}{1,.3,0.,0.1}

\numberwithin{equation}{section}
\renewcommand\theequation{\arabic{section}.\arabic{equation}}

\usepackage{tikz}

\usepackage{tcolorbox}

\newcommand{\lp }[0]{\left (}
\newcommand{\rp }[0]{\right )}
\newcommand{\llp }[0]{\left [}
\newcommand{\rrp }[0]{\right ]}

\newcommand{\med}[1]{\langle #1\rangle}

\newcommand{\vx}{\vec{x}}
\newcommand{\vy}{\vec{y}}

\def\TH{\text{\tiny th}}
\def\pk{\text{\tiny pk}}

\newcommand{\be}{\begin{equation}\begin{aligned}}
\newcommand{\ee}{\end{aligned}\end{equation}}

\newcommand{\bbe}{\begin{align}}
\newcommand{\eee}{\end{align}}

\newcommand{\bea}{\begin{eqnarray}}
\newcommand{\eea}{\end{eqnarray}}

\def\beq{\begin{equation}}
\def\eeq{\end{equation}}

\def\d{{\rm d}}

\def\beqa{\begin{eqnarray}}

	\def\eeqa{\end{eqnarray}}
	
\def\lsim{\mathrel{\rlap{\lower4pt\hbox{\hskip0.5pt$\sim$}}
		\raise1pt\hbox{$<$}}}         
\def\gsim{\mathrel{\rlap{\lower4pt\hbox{\hskip0.5pt$\sim$}}
		\raise1pt\hbox{$>$}}}         

\def\d{{\rm d}}

\def\pk{{\text{\tiny pk}}}

\def\d{{\rm d}}

\def\eeqa{\end{eqnarray}}

\def\bq{\begin{quote}}
\def\eq{\end{quote}}

 at 10truept
\newcommand{\arXiv}[2]{\href{http://arxiv.org/pdf/#1}{{\tt [#2/#1]}}}
\newcommand{\arXivold}[1]{\href{http://arxiv.org/pdf/#1}{{\tt [#1]}}}

\def\eeqa{\end{eqnarray}}
\def\lsim{\mathrel{\rlap{\lower4pt\hbox{\hskip0.5pt$\sim$}}
    \raise1pt\hbox{$<$}}}         
\def\gsim{\mathrel{\rlap{\lower4pt\hbox{\hskip0.5pt$\sim$}}
    \raise1pt\hbox{$>$}}}         

\def\G{\text{\tiny G}}
\def\NG{\text{\tiny NG}}

\title{ The  Ineludible non-Gaussianity of the Primordial Black Hole Abundance}
\author[a]{V. De Luca,}
\author[a]{G. Franciolini,}
\author[b]{A. Kehagias,}
\author[c,d]{M. Peloso,}
\author[a,e]{A.~Riotto,}
\author[f]{\textcolor{White}{}C. \"{U}nal}

\affiliation[a]{D\'epartement de Physique Th\'eorique and Centre for Astroparticle Physics (CAP), Universit\'e de Gen\`eve, 24 quai E. Ansermet, CH-1211 Geneva, Switzerland}
\affiliation[b]{Physics Division, National Technical University of Athens, 15780 Zografou Campus, Athens, Greece}
\affiliation[c]{Dipartimento di Fisica e Astronomia ``G. Galilei'', Universit\`a degli Studi di Padova, via Marzolo 8, I-35131, Padova, Italy}
\affiliation[d]{INFN, Sezione di Padova, via Marzolo 8, I-35131, Padova, Italy}
\affiliation[e]{CERN, Theoretical Physics Department, Geneva, Switzerland}
\affiliation[f]{CEICO, Institute of Physics of the Czech Academy of Sciences (CAS), Na Slovance 1999/2, 182 21 Prague, Czechia}

\abstract{We study the   formation of primordial black holes when they are generated  by the collapse of large overdensities in the early universe. Since the density contrast is related
to the comoving curvature perturbation by a nonlinear relation, the overdensity statistics is unavoidably non-Gaussian.  We show that the abundance of primordial black holes at formation may not be captured by a perturbative approach which retains the first few cumulants of the non-Gaussian probability distribution. 
We provide two techniques  to calculate the  non-Gaussian abundance of primordial black holes at formation, one based on peak theory and the other on threshold statistics. Our results show that  the  unavoidable non-Gaussian nature of the inhomogeneities in the energy density  makes it  harder  to generate PBHs. We provide simple
(semi-)analytical expressions to calculate the non-Gaussian abundances of the primordial black holes and show that for both narrow and broad power spectra the gaussian case from threshold statistics is reproduced by increasing the amplitude of the power spectrum by a factor ${\cal O}(2\div 3)$.
}
\emailAdd{valerio.deluca@unige.ch}
\emailAdd{gabriele.franciolini@unige.ch}
\emailAdd{kehagias@central.ntua.gr}
\emailAdd{marco.peloso@pd.infn.it}
\emailAdd{antonio.riotto@unige.ch}
\emailAdd{unal@fzu.cz}
\begin{document}

\maketitle
\flushbottom

\section{Introduction}
\renewcommand{\theequation}{1.\arabic{equation}}
\setcounter{equation}{0}
\noindent
Since the first  detection of gravitational  waves originated by  the merging of two $\sim 30 M_\odot$ black holes \cite{ligo}, the idea that Primordial Black Holes (PBHs)
might form a considerable fraction of the dark matter \cite{PBH1,PBH2,PBH3} has attracted again much interest \cite{kam} (see Ref. \cite{revPBH} for a recent review).
A  popular mechanism for the formation of PBHs is the   scenario in which PBHs are originated from the enhancement of the curvature power spectrum at a given short length scale  due to some  features  \cite{revPBH}. 
If the power spectrum of the curvature perturbation is enhanced during inflation  to values  $\sim 10^{-2}$ on small scales and subsequently transferred   to radiation during the reheating process, 
PBHs may form from sizeable  fluctuations   if the latter   overcome the counter effect  of the radiation pressure.

Since the  perturbation of  fixed comoving size does not  collapse till it re-enters  the cosmological horizon, the  size of a PBH at formation  is related to  the horizon length and its mass $M$  is approximately  the mass contained in such a horizon volume.   Fluctuations  collapse immediately after horizon re-entry to form  PBHs  if they are sizeable enough. We indicate  by  
$\delta$  the overdensity and by  $\sigma_\delta^2$ its variance
\be
\sigma^2_\delta =\int \frac{{\rm d}^3k}{(2\pi)^3}\,W^2(k,R_H)\,P_{\delta}(k),
\ee
where $P_\delta$ is the overdensity power spectrum, $R_H$ being   the comoving horizon length  $R_H=1/aH$,  $H$ is the Hubble rate and $a$ the scale factor. The quantity $W(k,R_H)$ is a  window function, for which we choose a top-hat in real space. Under the assumption that the density contrast is a linear quantity obeying gaussian statistics, threshold statistics (or Press-Schechter) predicts that  the primordial mass fraction $\beta(M)$ of the universe stored into PBHs at the formation time   is given by\footnote{In the literature sometimes  this expression may be multiplied by a factor of 2 to account for the cloud-in-cloud problem \cite{exc}. There seems to be no agreement if this factor should be included
for PBHs. Numerically it makes little difference.}
\be\label{press-s}
P_{\G}(\delta>\delta_{\rm c})=\beta(M)=\int_{\delta_{\rm c}} \frac{{\rm d}\delta}{\sqrt{2\pi}\,\sigma_\delta}e^{-\delta^2/2\sigma_\delta^2}.
\ee
Here $\delta_{\rm c}$ is the threshold for formation of the PBHs which quantifies how large the overdensity perturbations must be and depends on the shape of the power spectrum   \cite{haradath,mg,musco}.
By defining 
\be
\nu_{\rm c}=\frac{\delta_{\rm c}}{\sigma_\delta},
\ee
 the Gaussian mass fraction can be well approximated by ($\nu_{\rm c}\gsim 5$)
 \be
 \label{omega}
\beta^{\rm th}_\G\simeq \sqrt{\frac{1}{2\pi\nu_{\rm c}^2}}
e^{-\nu_{\rm c}^2/2}.
 \ee
This expression for the PBH mass fraction comes about when identifying the PBHs with regions whose overdensity is above a given threshold, hence the name of threshold statistics.

Alternatively, one can identify the PBHs with the local maxima of the overdensity, and one may use peak theory \cite{bbks} to compute their mass fraction. In such a case one has \cite{b}\footnote{We differ slightly from the corresponding expression in  Ref. \cite{b}. First by a factor of 3  to account for the fact that 
one counts the number density of peaks  at superhorizon scales, but the PBHs formed once the overdensity crosses the horizon at a slightly later time \cite{mg} (see also section 3). Secondly, by the fact that we define the mass going into PBH to be $M=(4\pi/3)\overline{\rho} R_H^3$, where $\overline{\rho}$ is the background radiation density. More importantly, we use here the definition (\ref{nupk}) for the critical value $\nu_\pk$. We will give more details in section 3.  At the gaussian level, peak theory gives a PBH abundance which is systematically larger than the one provided by the threshold statistics \cite{b}. }
\be
 \label{omegapk}
\beta^{\pk}_\G\simeq \frac{1}{3\pi}\left(\frac{\langle k^2\rangle}{3}\right)^{3/2}\,R_H^3\,(\nu_{\pk}^{2}-1)\,
e^{-\nu_{\pk}^2/2}\,\,\,\,{\rm with}\,\,\,\,\langle k^2\rangle=\frac{1}{\sigma_\delta^2}\int \frac{{\rm d}^3k}{(2\pi)^3}\,k^2\,P_{\delta}(k),
 \ee
 where now \cite{mg}
 \be
 \label{nupk}
 \nu_\pk=\frac{\delta_\pk^{\rm c}}{\sigma_\delta},
 \ee
  and $\delta_\pk^{\rm c}$ is to be identified with the critical value of the overdensity at the center of the peak above which
 an initial perturbation eventually collapses into a PBH \cite{mg,musco}. Notice that here  we follow Refs. \cite{mg,haradath} and do not introduce  a  window function   for the peak theory. Indeed, for the examples we will discuss  the window function   is not strictly necessary because they are characterised by a well-defined scale in momentum space and the  corresponding  distribution   is already smooth on length scales smaller than that characteristic scale.   Also, in the case of peak theory a typical length pops out automatically, that is  the scale  $R_*$. 
 
The  gaussian expressions (\ref{omega}) and  (\ref{omegapk}) make  already manifest the essence of the problem we are going to discuss in this paper. 
PBHs are generated  through   very  large, but rare  fluctuations. Therefore, their mass fraction at formation  is extremely sensitive to changes in the tail of the fluctuation distribution and therefore  to any possible  non-Gaussianity in the density contrast  \cite{s3,pina,byrnes,ngtwo,ng1,ng2,ng3,ng33,ng4,ng5,ng6,noi,noi3,g}. This implies that  non-Gaussianities need to be accounted for  as they can alter the initial mass fraction of PBHs in a dramatic way. 
For instance, the   presence of a  primordial local non-Gaussianity in the comoving curvature perturbation  can significantly alter the number density of PBHs  through   mode coupling  \cite{localNG,byrnes2,jmcpbhng1,jmcpbhng2,sasaking,caner}. 
   
In this paper  we  will be dealing with a source of  non-Gaussianity which is unavoidably generated by the non-linear relation among the overdensity $\delta(\vx,t)$ ($t$ is the cosmic time) and the
comoving curvature perturbation $\zeta(\vx)$.  It is important to stress that this non-linear relation makes the overdensity non-Gaussian even if the curvature perturbation is gaussian. In this sense, the non-Gaussianity we will discuss here is  ineludible.

Let us briefly discuss
where this non-linearity relation comes from. 
As we mentioned above, in the early radiation-dominated universe, the PBHs are generated when  highly overdense regions gravitationally collapse directly into a black hole. 
Before collapse, the comoving sizes of such  regions are   larger than the horizon length and the  separate universe approach can be applied \cite{harada}.
One   therefore expands at  leading order in spatial gradients of the various observables, e.g. the overdensity.  At this stage, the slicing and the threading of the spacetime manifold are to be fixed.  For instance, the so-called comoving  gauge seems appropriate as it has been adopted to perform numerical relativity simulations to describe  the formation of PBHs and to calculate the  threshold for PBH formation \cite{haradath}. 

In the comoving  slicing, the overdensity 
turns out to be \cite{harada} 
  \be
 \label{curv}
 \delta(\vec x,t)=-\frac{8}{9a^2H^2}e^{-5\zeta(\vec x)/2}\nabla^2e^{\zeta(\vec x)/2}= - \frac{4}{9} \frac{1}{a^2 H^2} e^{-2 \zeta(\vx)} \lp  \nabla^2\zeta(\vx) +\frac{1}{2} \partial_i \zeta(\vx) \partial^i \zeta(\vx) \rp .
 \ee
As the universe expands,  the overdensity  grows.  Regions where it  becomes of order unity eventually stop  expanding and collapse. This happens when the comoving scale of such a region becomes of the order of the  horizon scale. Even though the gradient expansion approximation  breaks down, it has been used to obtain   an acceptable criterion for the PBH  formation (that is to compute the overdensity threshold) and this approximation has been confirmed to hold by  nonlinear numerical studies \cite{revPBH,shibata}.  

The standard procedure in the literature is to expand the relation (\ref{curv}) to first-order in $\zeta$
\be
\label{first}
\delta(\vec x,t)=-\frac{4}{9a^2H^2}\nabla^2\zeta(\vec x)
\ee
and to relate the power spectrum of the overdensity to the one of the curvature perturbation by the relation 
\be
P_{\delta}(k,t)=\frac{16}{81}\frac{k^4}{a^4H^4}P_\zeta(k).
\ee
The question is to what extent this is a good approximation given the fact that even tiny  changes (percent level) in the square root of the overdensity variance are exponentially amplified in the
PBH mass fraction.  

To get the feelings of the numbers, let us roughly estimate the impact of the exponential $e^{-2\zeta(\vx)}$.   Calling $k_\star$ the typical momentum of the perturbation, from  Eq. (\ref{first})  we get
\be
\zeta\simeq \frac{9a^2H^2}{4k_\star^2}\delta\simeq  \frac{9a^2H^2}{4k_\star^2}\delta_{\rm c}\simeq 0.15,
\ee
where we have taken the threshold $\delta_{\rm c}\simeq 0.5$ and $k_\star\simeq 2.7 a H$ \cite{mg}. This gives $e^{-2\zeta(\vx)}\simeq 0.7$. This looks as a small change, but in fact it  has 
an exponentially large  effect in the mass fraction when the corresponding overdensity variance is calculated.

The goal of this paper is to  deal with the  intrinsically non-Gaussian nature of the overdensity onto the mass fraction of PBHs. First of all, we will provide a simple argument
to convince the reader that the non-Gaussianity  introduced by the non-linear relation (\ref{curv}) between the overdensity and the gaussian curvature perturbation has an impact on the PBH mass fraction which may not be accounted for by a perturbative approach.  Based on this finding, we will proceed by computing the mass fraction taking into account such intrinsic non-Gaussianity. We will do so by using
two methods. 

Since PBHs may be thought to originate from peaks, that is, from maxima of the local overdensity, we will resort to peak theory  \cite{bbks} to calculate the probability of formation of the PBHs.  This method is based on the fact that for high values of the overdensity at the peaks, their location can be confused  with  the location of the peaks in the comoving curvature perturbation as long as such peaks are sufficiently spiky, that is if their curvature (proportional to the second spatial derivatives) is large enough at the center of the peak \cite{haradath}.

Alternatively, we will use the non-Gaussian threshold statistics and provide an exact expression for the probability to form PBHs. Both methods  indicate that the inevitable non-Gaussian nature of the overdensity makes more difficult to generate PBHs, independently from the shape of the power spectrum.

Let us also add a cautionary note. The intrinsic non-Gaussianity of the overdensity changes also the shape of the profile of the peaks which eventually give rise to PBHs upon collapse. Since the threshold 
depends on the shape of the overdensity, such non-Gaussianity influences as well the threshold value.
This will be discussed in a separate publication \cite{kmr}.

The paper is organised as follows. In section 2 we offer a  simple criterion to show that the intrinsic non-Gaussianity cannot be described by perturbative methods. Sections 3 and 4
will describe the two methods mentioned above. Section 5 contains our conclusions. The paper contains as well several appendices for the technical details.

\section{A simple criterion to show that intrinsic non-Gaussianity matters}
\renewcommand{\theequation}{2.\arabic{equation}}
\setcounter{equation}{0}
\noindent
In order to establish if the intrinsic non-Gaussianity introduced by the non-linear relation (\ref{curv}) is relevant, we start from the non-Gaussian threshold statistics developed in Ref. \cite{blm} and refined in Ref. \cite{noi}
by means of a  path-integral approach. We do not report all the details here and the  interested reader is refereed to those references for more details. We do not use here the window function which would introduce painful, but useless technicalities without changing the conclusions. Suffice to say that 
the probability of having the overdensity larger than a given threshold can be viewed as the one-point function of the  threshold quantity
\be
P(\delta>\delta_{\rm c})=\Big<\Theta(\delta-\nu_{\rm c}\sigma_{\delta})\Big>=\int [D\delta(\vx)]P[\delta(\vx)]
\Theta\Big(\delta(\vx)-\nu_{\rm c}\sigma_\delta\Big),
\ee
where $\Theta(x)$ is the Heaviside function.
By defining the connected correlators of the overdensity as 
\begin{eqnarray}
\Big<\delta(\vx_1)\cdots\delta(\vx_n)\Big>_c= \xi_n(\vx_1,\cdots,\vx_n),
\end{eqnarray}
one finds that, in the limit of large $\nu_{\rm c}$, the  threshold statistics is given by \cite{blm,noi}
\be
\label{final}
P(\delta>\delta_{\rm c})=\beta(M)=\frac{1}{\sqrt{2\pi \nu_{\rm c}^2}}\exp\left\{-\nu_{\rm c}^2/2+\sum_{n=3}^\infty
\frac{(-1)^{n}}{n!}\xi_{n}(0) \, (\delta_{\rm c}/\sigma_{\delta}^2)^n\right\},
\ee
where the label 0 means that the correlators are computed at equal points.
To see under which circumstances  the non-Gaussianity of the overdensity alters the predictions of the gaussian primordial abundance of PBHs in a significant way, we define dimensionless quantities, the cumulants, by  the relations
\be
S_n=\frac{\xi_n(0)}{\left(\xi_2(0)\right)^{n-1}}=\frac{\overset{n-{\rm times}}{\langle\overbrace{\delta(\vx)\cdots\delta(\vx)}\rangle}_c }{\sigma^{2(n-1)}_{\delta}}.
\ee
Following Ref. \cite{noi} we may  define the fine-tuning $\Delta_n$  to be the response of the PBH abundance to the 
introduction of the $n$-th cumulant as 
\be
\Delta_n=\frac{\d\ln \beta(M)}{\d\ln S_n}.
\ee
Each cumulant allows to express the   non-Gaussian PBH abundance  in terms 
of    the gaussian abundance as
\begin{eqnarray}
 \frac{\beta^\TH_{\NG}(M)}{\beta^\TH_{\G}(M)}=e^{\Delta_n}.
 \end{eqnarray}
 This implies that the PBH abundance is exponentially sensitive to the non-Gaussianity unless $\Delta_n$ is  in absolute value smaller than unity
 \be
|\Delta_n|\lsim 1.
\ee
Inspecting  Eqs. (\ref{omega}) and (\ref{final}), we see that
\be
| \Delta_n|=\frac{1}{n!}\left(\frac{\delta_{\rm c}}{\sigma_{\delta}}\right)^2|S_n|\delta_{\rm c}^{n-2}.
\ee
This tells us that intrinsic non-Gaussianity in the overdensity alters exponentially the gaussian prediction for the PBH abundance unless
\be
\label{crit}
\left|S_n\right|\lsim \left(\frac{\sigma_{\delta}}{\delta_{\rm c}}\right)^2\frac{n!}{\delta_{\rm c}^{n-2}}.
\ee
To investigate how restrictive this condition is, we take the  simplest  case possible, i.e. a very narrow power spectrum for the comoving curvature perturbation which we approximate by a Dirac delta
\be
P_\zeta(k)=\frac{2\pi^2}{k^3}\mathcal{P}_{\zeta}(k)\,\,\,\,{\rm and}\,\,\,\, \mathcal{P}_{\zeta}(k) = A_s k_\star\delta_D(k-k_\star).
\label{Pzeta-delta} 
\ee 
Here $A_s$ is the amplitude of the power spectrum and $k_\star$ is the characteristic scale of the power spectrum. Its relation with the cosmological horizon at formation $R_H$  has to be fixed running numerical simulations \cite{mg,musco}. For the case at hand, it is given by $k_\star\simeq 2.7/R_H$ (more comments on this later on). We do not report all the technical details here, which can be found in Appendix A, where we have consistently calculated the variance, the skewness $S_3$ and the kurtosis $S_4$ up to third-order in perturbation theory (in the power spectrum $P_\zeta$, that is up to $A_s^3$). We get
\begin{eqnarray}
\langle\delta^2\rangle_c&=&\sigma_\delta^2 
=c_\star^2 k_\star^4 A_s \lp 1 + \frac{133}{6} A_s + \frac{511}{3} A_s^2\rp,\nonumber\\
\langle \delta^3  \rangle_c 
&=&  -c_\star^3 k_\star^6 12 A_s^2  \lp 1  + \frac{3889}{108} A_s\rp,\nonumber\\
\langle \delta^4  \rangle_c 
&=&
 240 c_\star^4 k_\star^8 A_s^3,\nonumber\\
c_\star k_\star^2&=&\frac{4}{9} \lp \frac{k_\star}{a H}\rp ^2   \simeq 3.2
.
\label{d234c-res}
\end{eqnarray}
One can check that the  criterion (\ref{crit}) for the skewness (kurtosis) gives the lower bound 
\be\label{bound}
A_s\gsim 6.0 \,(4.0)\cdot 10^{-3},
\ee
where we have taken $\delta_{\rm c}=0.5$. We now impose the condition that the PBHs form at most the totality of dark matter, which provides an upper bound on their  mass fraction given by 
\be
\beta\lsim 1.3 \times 10^{-9} \lp \frac{M}{M_\odot}\rp ^{1/2}.
\ee
For instance, for PBH masses around the interesting value of   $10^{-12} M_\odot$ \cite{noi1,noi2}, one would get from the gaussian mass fraction (\ref{omega}) $\beta\sim  10^{-15}$, $\nu_{\rm c}\simeq 8$ and therefore $A_s\simeq 3.7 \cdot 10^{-4}$. This figure violates the bound required \eqref{bound} to neglect the non-Gaussianity by one order of magnitude.  
More importantly, the kurtosis does not provide a bound which is much weaker than the skewness. This signals the breaking of the perturbative approach and calls for a more refined treatment. 

The same conclusion can be obtained  in the case where the power spectrum of the comoving curvature perturbation is parametrised by a log-normal shape of the form
\be
\mathcal{P}_\zeta (k) = \frac{A_g}{\sqrt{2\pi}\sigma}
\exp \llp -\frac{ \ln ^2\lp k/ k_\star \rp }{2 \sigma^2 }\rrp.
\ee
Using the results in Appendix A, one finds the following (for $\sigma=0.2$) 
\begin{eqnarray}
\langle\delta^2\rangle_c&=&\sigma_\delta^2 
=1.4 \cdot c_\star^2 k_\star^4 A_g \lp 1  + 20   A_g + 150 A_g^2\rp,\nonumber\\
\langle \delta^3  \rangle_c 
&=&  - 18 \cdot  c_\star^3 k_\star^6  A_g^2  \lp 1  + 34 A_g\rp,\nonumber\\
\langle \delta^4  \rangle_c 
&=&
400 \cdot c_\star^4 k_\star^8 A_g^3.
\label{d234c-gauss}
\end{eqnarray}
The criterion (\ref{crit}) in this case results in a lower bound 
\be
A_g\gsim 3.8 \,(2.2)\cdot 10^{-3},
\ee
for the skewness and kurtosis respectively, while requiring again $\beta\sim 10^{-15}$ for $M\sim 10^{-12} M_\odot$ gives $A_g=2.5\cdot 10^{-4}$. Again we do not see signs of convergence in the perturbative approach.

\section{The non-Gaussian probability from peak theory}\label{id-peaks}
\renewcommand{\theequation}{3.\arabic{equation}}
\setcounter{equation}{0}
\noindent
Having shown that perturbation theory fails to provide the probability for PBH formation, we first resort to peak theory \cite{bbks}.
As we already mentioned in the introduction,  PBHs trace the peaks of the radiation density field  on superhorizon scales where the number of peaks per comoving volume is constant.  Notice that we are dealing with peaks of the overdensity rather than the peaks of the curvature perturbation. This is because one cannot impose any constraint on the value of the gravitational potential (or curvature perturbation) on superhorizon scales  because constant gravitational potentials  cannot   lead to any observable effect. 
Nevertheless, one can start from the following important point:  large threshold   peaks of  the overdensity may be identified within a Hubble volume with  the peaks of the curvature perturbation if the  Laplacian of the   curvature perturbation (that is the curvature of the peak) at the peak is large enough \cite{haradath}. More in details, one can show that if the value of $\delta$ is comparable to the threshold value  at a peak, one can   find the associated   peak of $\zeta$ well inside the horizon patch and centered at the peak of $\delta$ as long as the peaks in $\zeta$ is spiky enough. Let us elaborate about this point in the next subsection.

\subsection{Spiky peaks of the curvature perturbation may be confused with peaks of the overdensity for large thresholds}
\noindent
The argument given in Ref. \cite{haradath} is as follows. 
Let us consider the nonlinear expression (\ref{curv}) relating $\delta$ and $\zeta$ on superhorizon scales and in radiation domination  
\be
\label{delta}
\delta(\vx,t)=-\frac{4}{9a^2 H^2}e^{-2\zeta(\vx)}\left[\nabla^2\zeta(\vx)+\frac{1}{2}\partial_i\zeta(\vx)\partial^i\zeta(\vx)\right].
\ee 
We can expand the comoving curvature perturbation $\zeta(\vx)$ for points $\vx$ around the peak position $\vx_\pk$ of the overdensity \footnote{We indicate
by $\partial_i\zeta(\vx_\pk)$ the gradient $\partial_i\zeta(\vx)$ computed at $\vx_\pk$, and so on.}  $\delta(\vx,t)$ 
\be
\label{zeta}
\zeta(\vx)=\zeta(\vx_\pk)+\partial_i\zeta(\vx_\pk)(x^i-x^i_\pk)+\frac{1}{2}\partial_i\partial_j\zeta(\vx_\pk)(x^i-x^i_\pk)(x^j-x^j_\pk).
\ee
Around such a peak 
we can also write
\begin{eqnarray}
\label{delta1}
\delta(\vx_\pk,t)&\simeq& -\frac{4}{9a^2 H^2}e^{-2\zeta(\vx_\pk)}\nabla^2\zeta(\vx_\pk),
\end{eqnarray}
where we neglected the second term in the square bracket since 
its contribution is of higher order in $\zeta$ with respect to \eqref{delta1}.

Since the peak amplitude of the overdensity must be  larger than some critical  value $\delta^{\rm c}_\pk$, we deduce that the curvature of the peak in $\zeta$ is bounded from above
\be
\label{c}
-\nabla^2\zeta(\vx_\pk)> \frac{9a^2H^2}{4}e^{2\zeta(\vx_\pk)}\delta^{\rm c}_\pk.
\ee
This is what we meant by saying that the peaks in $\zeta$ must be spiky enough.
Now, the  peak in $\zeta$ is located in $\vy_\pk$ such that $\partial_i\zeta(\vy_\pk)=0$, or 
\be
\partial_i\zeta(\vx_\pk)+\partial_{i}\partial_{j}\zeta(\vx_\pk)(y_\pk^i-x^i_\pk)=0\,\,\,\,{\rm or}\,\,\,\,(y_\pk^i-x^i_\pk)=-(\zeta^{-1})^i_{\,j}(\vx_\pk)\partial_j \zeta(\vx_\pk),
\ee
where we have used in the last passage the notation $\partial_i\partial_j\zeta(\vx_\pk)=\zeta_{ij}(\vx_\pk)$.
Performing a rotation of the coordinate axes to be aligned with the principal axes  of the constant-$\zeta$ ellipsoids gives the eigenvalues of the
shear tensor $\zeta_{ij}$ to be equal to $-\sigma_2\lambda_i$, where 
 $\sigma_{2}$ is the characteristic root-mean-square variance of the components of $\zeta_{ij}$ (that of $\partial_i\zeta$ is $\sigma_1$) and 
 \be
 \lambda_i\simeq\frac{\gamma\nu}{3}, \,\,\,\,
\nu=\frac{\zeta(\vx_\pk)}{\sigma_0},\,\,\,\,\, \gamma=\frac{\sigma_1^2}{\sigma_0\sigma_2}\,\,\,\,{\rm and}\,\,\,\,\sigma_j^2=\int \frac{k^2{\rm d}k}{2\pi^2}P_\zeta(k) k^{2j}=
\int \frac{{\rm d}k}{k}{\cal P}_\zeta(k) k^{2j}.
\ee
 The crucial point is now that the moments $\sigma_j^2$ are typically much smaller than $(aH)^j$ (because of the presence of the amplitude of the power spectrum). From Eq. (\ref{c}), we deduce that
 \be\label{spiky-e}
 -\nabla^2\zeta(\vx_\pk)\sim \lambda_i\sigma_2 > \frac{9a^2H^2}{4}e^{2\zeta(\vx_\pk)}\delta^{\rm c}_\pk\gg \sigma_2
 \ee
 and therefore $\lambda_i\sim\gamma\nu\gg 1$ (the probability to have negative eigenvalues is small for large curvatures around the peak \cite{bbks}) . This implies
 \be
 |y_\pk^i-x^i_\pk|\simeq |\sigma_1/\sigma_2\lambda_i|\ll  |\sigma_1/\sigma_2|\lsim 1/aH,
 \ee
 where in the last equality we have used the fact that $\sigma_1/\sigma_2\simeq k_\star^{-1}\lsim R_H$.
 Therefore the high overdensity peaks  in $\delta$ lie close to the peaks of the curvature perturbation (i.e. within the Hubble volume) if the latter are characterised by a large second derivatives at the origin of the peak. This statement if of course valid in  the probabilistic sense.

 Since some approximations have been made along the way, in Appendix B the reader can find a  numerical simulation we have performed to  support this  result. 
 
 \subsection{The calculation of the  probability from peak theory}
  \noindent
If the argument above is correct, one can associate the number of rare peaks in the overdensity with the  number of peaks in the curvature perturbation which are spiky enough, see Eq.~\eqref{spiky-e}.   Therefore, 
expanding around the peak location $\vx_\pk$  of $\zeta$ (where $\partial_i\zeta(\vx_\pk)=0$) we can write 
\begin{eqnarray}
\delta(\vx_\pk,t)&=&-\frac{4}{9a^2 H^2}e^{-2\zeta(\vx_\pk)}\left[\nabla^2\zeta(\vx_\pk)+\frac{1}{2}\partial_i\zeta(\vx_\pk)\partial^i\zeta(\vx_\pk)\right]=
- \frac{4}{9a^2 H^2}e^{-2\zeta(\vx_\pk)}\nabla^2\zeta(\vx_\pk)\nonumber\\
&=& \frac{4}{9a^2 H^2}e^{-2\sigma_0\nu}x\sigma_2,
\end{eqnarray}
where 
\be
\nu=\frac{\zeta(\vx_\pk)}{\sigma_0}\,\,\,\,{\rm and}\,\,\,\, x=-\frac{\nabla^2\zeta(\vx_\pk)}{\sigma_2}.
\ee 
Since the number of peaks (if spiky enough) in $\zeta$ is approximately the number of peaks in $\delta$,  we can use the expression (A.14) of Ref. \cite{bbks} to find the number of peaks of the overdensity
\be
{\cal N}_\pk(\nu,x){\rm d}\nu{\rm d}x=\frac{e^{-\nu^2/2}}{(2\pi)^2R_*^3}f(x)\frac{{\rm exp}[-(x-x_*)^2/2(1-\gamma^2)]
}{[2\pi(1-\gamma^2)]^{1/2}}{\rm d}\nu{\rm d}x,
\ee
where
\be
R_*=\sqrt{3}\frac{\sigma_1}{\sigma_2},\,\,\,\,\gamma=\frac{\sigma_1^2}{\sigma_0\sigma_2},\,\,\,\,{\rm and}\,\,\,\, x_*=\gamma\nu,
\ee
and $f(x)$ is provided by the expression
\be
f(x) = \frac{(x^3-3x)}{2}\bigg[{\rm erf}\left(x\sqrt{\frac{5}{2}}\right) + {\rm erf}\left(\frac{x}{2}\sqrt{\frac{5}{2}}\right)\bigg] + \sqrt{\frac{2}{5\pi}}\bigg[\left(\frac{31x^2}{4} + \frac{8}{5}\right)e^{-\frac{5x^2}{8}}+ \left(\frac{x^2}{2}-\frac{8}{5}\right)e^{-\frac{5x^2}{2}}\bigg].
\ee
Thus the number density of non-Gaussian peaks of the overdensity above a given threshold $\delta_{\pk}^{\rm c}$ is simply given by 
\begin{tcolorbox}[colframe=white,arc=0pt]
\vspace{-.15cm}
\begin{eqnarray}
\label{Npeakformula}
{\cal N}_\pk=\int_{-\infty}^\infty\d\nu\int_{x_\delta^{\rm c}(\nu)}^\infty\d x\,\frac{e^{-\nu^2/2}}{(2\pi)^2R_*^3}f(x)\frac{{\rm exp}[-(x-x_*)^2/2(1-\gamma^2)]
}{[2\pi(1-\gamma^2)]^{1/2}},\nonumber\\
&&
\end{eqnarray}
\end{tcolorbox}
\noindent
where 
\be\label{cond}
x_\delta^{\rm c}(\nu)\simeq\frac{9a^2H^2}{4\sigma_2}e^{2\sigma_0\nu}\delta^{\rm c}_\pk
\ee
accounts for the fact that only large enough Laplacian values at the peak of the curvature perturbation have to be accounted for, see Eq. (\ref{c}). Notice that 
if we take the lower limit (\ref{cond}) at $\nu=0$, $x_\delta^{\rm c}(0)\simeq(9a^2H^2/4\sigma_2)\delta^{\rm c}_\pk$, we automatically reproduce the gaussian case. We have checked numerically that in such a case, the peak theory abundance of PBHs obtained from the number density (\ref{Npeakformula})  with $x_\delta^{\rm c}(0)$ reproduces the abundance (\ref{omegapk}) within a factor of order unity. This gives us extra confidence that identifying large threshold peaks in $\delta$ with the spiky enough peaks in $\zeta$ is a correct
procedure.
From the expression above one can see that the narrower is the power spectrum (that is the closer to unity is the parameter $\gamma$) the more the integrand is peaked at the value $x\simeq x_*\simeq \nu$. 

We  conclude that the non-liner relation between the curvature perturbation and the overdensity makes it harder to generate PBHs, independently from the shape of the curvature perturbation power spectrum.

From  the knowledge of the  number density of peaks  ${\cal N}_\pk$ 
we can compute the mass fraction of PBHs  $\beta$ at the time of the formation $t_{\rm f}$. 
Since PBHs trace the peaks of the radiation density field  on superhorizon scales and since  the number of peaks per comoving volume is constant,
the number of enough sizeable peaks on superhorizon scales provides the number of PBHs formed once the overdensity has crossed  the horizon
and one has  properly rescaled it to  the formation time \cite{mg,musco}. 

The next question is therefore what defines the horizon crossing.
In cosmology we are used to the concept of the horizon crossing associated to a given comoving wavelength $k^{-1}$ and we say that horizon crossing takes place when $k=aH$.  In the case of PBHs,  the large inhomogeneities have characteristic profiles in coordinate space and therefore it is not immediate to associate to them a given wavelength or momentum.
The procedure we will follow is the one adopted to define the threshold for collapse \cite{musco}. Suppose the overdensity has an average  profile in real space given by \cite{bbks}\footnote{As mentioned already in the introduction, we do not include here the non-Gaussianities in the average overdensity profile, whose effect we will study elsewhere \cite{kmr}. As for the variance around the average profile, it is negligible for  $\delta_\pk^{\rm c}/\sigma_\delta\gg 1$.}
\be
\label{profile}
\overline{\delta}(r,t)=\delta_\pk\frac{\xi_2(r,t)}{\sigma^2_\delta(t)},
\ee
where $\xi_2(r,t)$ is the two-point correlator.
One can define a scale $r_m$ through  the relation
\be
r_m^3 = \frac{\int_0^{r_m}\d r\,\overline{\delta}(r,t) r^2}{\overline{\delta}(r_m,t)}.
\ee
 This scale is relevant since one can show that the threshold for PBH formation is given by \cite{mg,musco}
 \be
 \delta_\pk^{\rm c}=\frac{\delta_{\rm c}}{3}\frac{\sigma^2_\delta(t_m)}{\xi_2(r_m,t_m)},
 \ee
 where $\delta_{\rm c}=3\overline{\delta}(r_m,t_m)$, 
 since $r_m$ is precisely  the scale at which the compaction function ${\cal C}\simeq 2\delta M/ar$ (being $\delta M$ the overmass generated by the averaged curvature perturbation) is maximised \cite{musco}. 
Such a maximum is located  at distances larger than the cosmological horizon. 
It is then natural to define the ``horizon crossing" as the time at which\footnote{The condition should read $e^{\zeta(r_m)}a(t_m)H(t_m) r_m=1$, but $\zeta(r_m)\ll \zeta(x_\pk)$ and we can safely neglect this correction.}  $a(t_m)H(t_m) r_m=1$. Numerical simulations must  provide a relation between the scale $r_m$ and the characteristic momentum
 appearing in the power spectrum of the curvature perturbation. 

The mass fraction at formation time (that is when the horizon forms) from peak theory will then be 
\be
\beta^\pk_\NG =  \frac{M(R_H)}{\overline{\rho}_{\rm f}}\frac{a_m^3}{a_f^3} {\cal N}_\pk,
\ee
where $M(R_H)$ is the mass of the PBH associated with the horizon size \footnote{In case, one can take into account that  the PBH mass is not precisely the expression $M(R_H)$, but scales
with the  initial perturbations \cite{j}.} $R_H$, 
\be
M(R_H) = \frac{4\pi}{3}\overline{\rho}_m R_H^3(t_m),
\ee
  and $\overline{\rho}_{\rm f}$ and  $\overline{\rho}_m$ are the background radiation energy densities at the time of
formation and horizon crossing, respectively. Numerical simulations show that the ratio $a_f/a_m$ is rather independent from the shape of the power spectrum and $\sim 3$ \cite{mg}. We therefore have
\be
\beta^\pk_\NG \simeq 3\cdot\frac{4\pi}{3}  R_H^3 {\cal N}_\pk.
\ee

\subsection{The log-normal power spectrum}
We assume a power spectrum of the form
\be\label{PS}
\mathcal{P}_\zeta(k) =\frac{A_g}{\sqrt{2 \pi} \sigma} \exp \llp -\frac{ \ln ^2\lp k/ k_\star \rp }{2 \sigma^2 }\rrp,
\ee
where changing the value of  $\sigma$ changes the broadness of the power spectrum.
 For the case at hand it turns out that  \cite{mg}
\be
 a_m H_m =\frac{1}{R_H}= \frac{1}{2.7}k_\star
\ee
and one has to choose the critical value $\delta_{\pk}^{\rm c}=1.16$  corresponding to $\delta_{\rm c} = 0.51$ \cite{mg,musco}.

\begin{figure}[t]
\includegraphics[width = 0.6\linewidth]{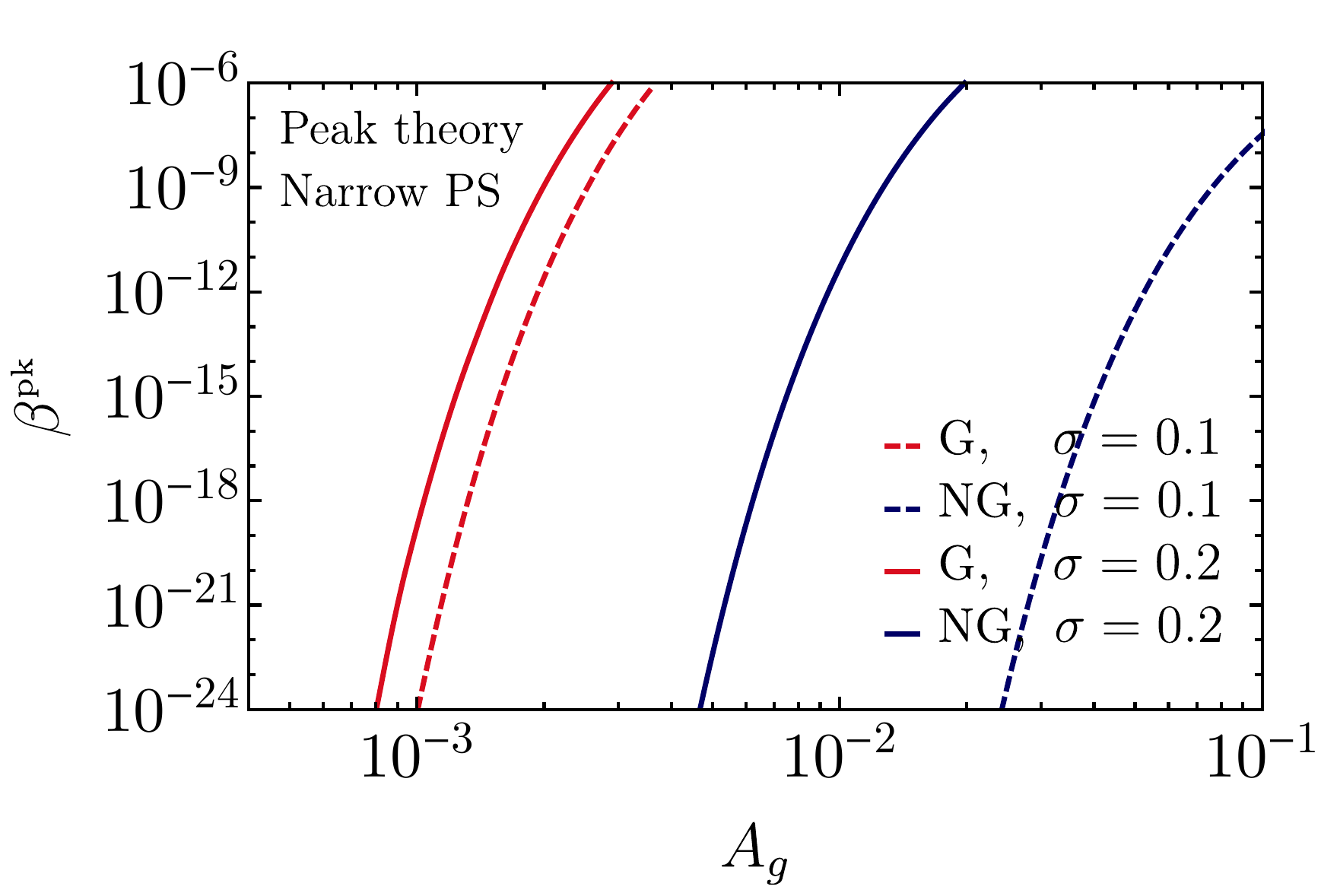}
\centering
\caption{\it
Mass fraction $\beta^{\text{{\tiny {\rm pk}}}}$ as a function of $A_g$ for log-normal power spectrum (PS)  computed using peak theory for both the gaussian and the non-Gaussian case.}
 \label{fig-betath}
\end{figure}

In Fig.~\ref{fig-betath} we plot the mass fraction  for various values of $\sigma$ as a function of $A_g$. 
We see that the inclusion of the intrinsic non-Gaussian effects systematically lowers the PBH abundance (having kept fixed the amplitude of the power spectrum of the curvature
perturbation). Said  in other words, keeping the amplitude of the fluctuations fixed, it is  more difficult to generate PBHs. This will remain true also using the threshold statistics, as we show in the next Section.
Quantitatively, in the case considered, for the usual value of $\beta \sim 10^{-15}$ necessary for PBHs to be all the dark matter in the universe for masses of the order of $10^{-12}\,M_\odot$, we find that in the gaussian case the value of the amplitude is consistent with the one reported in Ref.  \cite{mg} once the difference in the normalisation of the power spectrum is taken into account, while the non-Gaussian abundance is suppressed.

\subsection{Broad power spectrum}
We also  consider a broad  power spectrum, that is a top-hat function with amplitude $A_t $ as
\be
\mathcal{P}_\zeta (k) = A_t\, \Theta (k_{\rm max}-k)\, \Theta (k-k_{\rm min}) 
\ee
where $\Theta$ stands again for the Heaviside step function and $k_{\rm max}\gg k_{\rm min}$,  such that the scale $k_{\rm min}$ in practice does not participate in the PBH formation \cite{mg}.  In this case one finds $k_{\rm max}\simeq 3.5/r_m$,  $\delta_{\rm c}$ is again 0.51, and  $\delta_\pk^{\rm c}\simeq 1.22$ \cite{mg} and the variances are obtained by putting $a_mH_m$ as the  infrared cut-off since the unphysical long wavelength modes should be disregarded.
\begin{figure}[t]
\includegraphics[width = 0.6\linewidth]{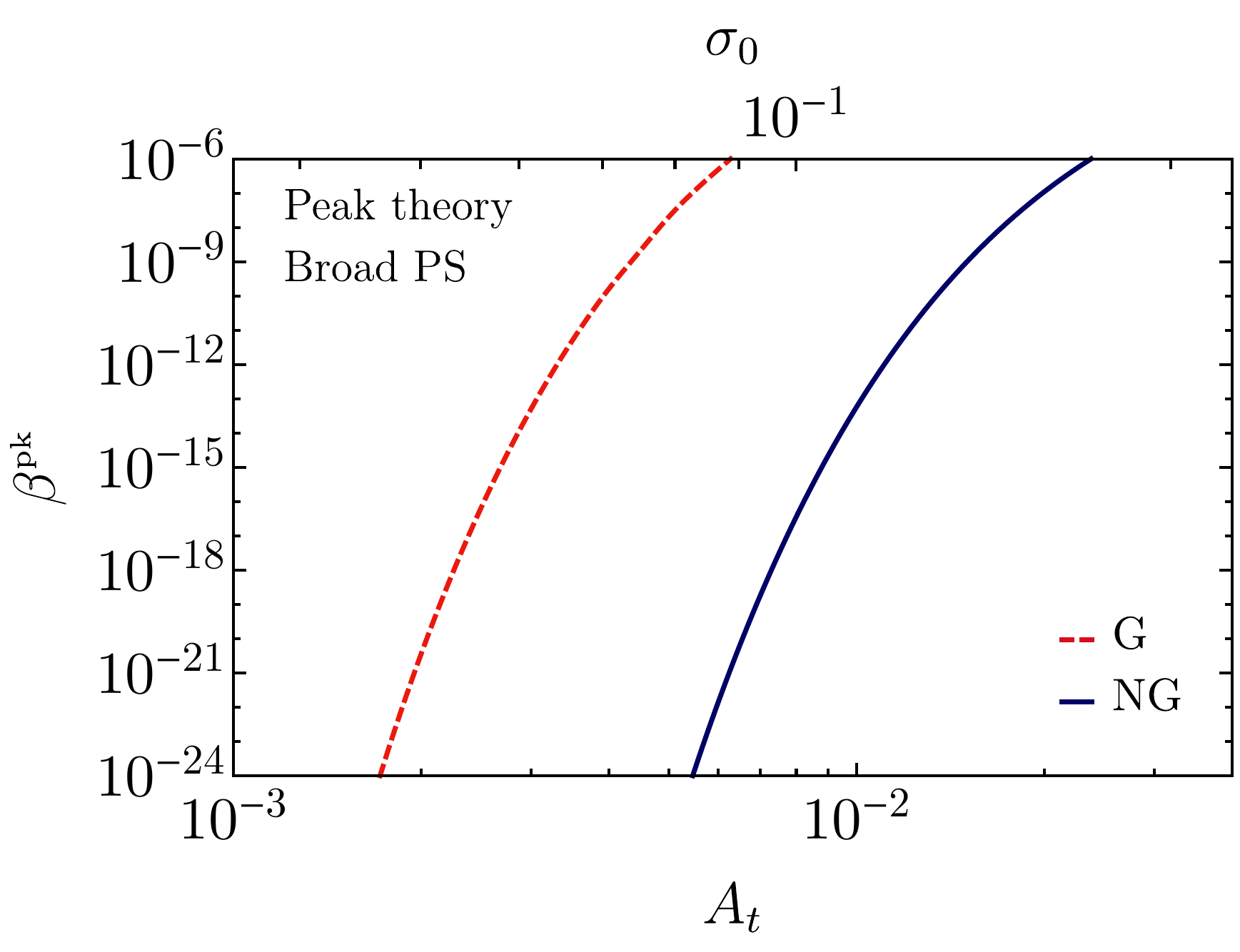}
\centering
\caption{\it
Mass fraction $\beta^{\text{{\tiny {\rm pk}}}}$ as a function of $A_t$ for the broad (top-hat) power spectrum  computed using peak theory for both the gaussian and the non-Gaussian case. In this plot (and in the following) we show in the horizontal axes the value of the amplitude of the power spectrum and its corresponding root of the variance $\sigma_0$. 
}
 \label{fig-betapk-2}
\end{figure}
Fig.~\ref{fig-betapk-2} shows the mass fraction as a function of $A_t$.\footnote{We do not introduce a window function to be able to compare with the gaussian results of Ref. \cite{mg} which are reproduced in the gaussian case.}
As predicted, both for narrow spectra and broad ones,  the intrinsic non-Gaussianity in the overdensity makes it harder to produce PBHs.

\section{The non-Gaussian probability from  threshold statistics}
\renewcommand{\theequation}{4.\arabic{equation}}
\setcounter{equation}{0}
\noindent
In this section we present an alternative way to calculate the non-Gaussian probability to form PBHs which does not rely on the fact that spiky peaks of the curvature perturbation coincide with peaks of the overdensity for large thresholds.   The price to pay is that we will be dealing with the threshold statistics (the threshold being   identified with $\delta_{\rm c}$ \cite{mg}). This might be not a great sacrifice as regions characterised by large thresholds are likely to be regions of maxima of the overdensity \cite{hof}. The gain is that  the expressions we are going to obtain are exact.

Let us consider again the curvature perturbation $ \zeta(\vx)$ as a random field. Following the notation of the Appendix A of Ref. \cite{bbks}, we define
\be
\zeta_i = \partial_i \zeta, \quad  \zeta_{ij} = \partial_i \partial_j \zeta.
\ee
The correlations of these fields are provided by the expressions
\begin{eqnarray}
\med{\zeta \zeta} &=& \sigma_0^2, \\
\med{\zeta \zeta_{ij}} &=& -\frac{\sigma_1^2}{3}\delta_{ij}, \\
\med{\zeta \zeta_{i}} &=& 0, \\
\med{\zeta_i \zeta_i} &=& \frac{\sigma_1^2}{3}\delta_{ij}, \\
\med{\zeta_{ij} \zeta_{kl}} &=& \frac{\sigma_2^2}{15}(\delta_{ij}\delta_{kl} + \delta_{ik}\delta_{jl}+ \delta_{il}\delta_{jk}), \\
\med{\zeta_i \zeta_{jk}} &=& 0.
\end{eqnarray}
These variances will be computed numerically using the Fourier transform of the top-hat window function in real space, that is

\be
\sigma_j^2=\int \frac{k^2{\rm d}k}{2\pi^2}W^2(k,r_m) P_\zeta(k) k^{2j},\,\,\,\,W(k,r_m)=3\frac{\sin(k r_m)-k r_m \cos(k r_m)}{(k r_m)^3}.
\ee
The matrix $-\zeta_{ij}$ can be diagonalized with eigenvalues $\sigma_2 \lambda_i$, ordered such that $\lambda_1 \geq \lambda_2 \geq \lambda_3$.
Thus we define
\be
 x = - \frac{\nabla^2 \zeta}{\sigma_2} = \lambda_1 + \lambda_2 + \lambda_3, \quad  y = \frac{\lambda_1-\lambda_3}{2}, \quad  z = \frac{\lambda_1-2\lambda_2+\lambda_3}{2}.
\ee
Introducing again $\nu = \zeta(\vx)/\sigma_0$, the correlations become
\be
\med{\nu^2} = 1, \quad \med{x^2} = 1, \quad \med{x\nu} = \gamma, \quad \med{y^2} = 1/15, \quad \med{z^2} = 1/5
\ee
and all the others are zero.
The joint gaussian probability distribution for these variables is provided by the expression (from now on we will label $\eta_i \equiv \zeta_i$)
\be
P(\nu, \vec{\eta} ,x,y,z) \d \nu \d^3 \eta \d x\d y \d z = N |2y(y^2-z^2)|e^{-Q} \d \nu \d x \d y \d z \frac{\d ^3 \eta}{\sigma_0^3}
\ee
as a function of 
\be
2Q = \nu^2 + \frac{(x-x_*)^2}{(1-\gamma^2)} + 15y^2 + 5z^2 + \frac{3 \vec{\eta} \cdot \vec{\eta}}{\sigma_1^2} 
\ee
and
\be
x_*= \gamma \nu, \quad \gamma = \frac{\sigma_1^2}{\sigma_0 \sigma_2}, \quad 
N = \frac{(15)^{5/2}}{32\pi^3}\frac{6 \sigma_0^3}{\sigma_1^3(1-\gamma^2)^{1/2}}.
\ee
The variables $y$ and  $z$ are unconstrained and we  integrate  them out. With the ordering of the eigenvalues previously defined, we see that the variable $z$ lies in the range $[-y,y]$, while $y \geq 0$. 
The result is therefore given by\footnote{
Notice that, assuming the linear relation between $\delta$ and $\zeta$ as in Eq.~\eqref{first}, one recovers the Press-Schechter result in Eq.~\eqref{press-s}.}
\be
\label{probnoch}
P(\nu, \vec{\eta} ,x) \d \nu \d^3 \eta \d x = C e^{-Q_2}\d \nu \d^3 \eta \d x,
\ee
where we have defined
\be
C = \frac{6 \sqrt{3}}{8\pi^{5/2}\sqrt{2(1-\gamma^2)}\sigma_1^3}
\ee
and
\be
\label{Q2}
2Q_2 = \nu^2+ \frac{(x-x_*)^2}{(1-\gamma^2)} +  \frac{3 \vec{\eta} \cdot \vec{\eta}}{\sigma_1^2}.
\ee
We can then write the $\delta$ as a function of these variables as
\be
\delta(\vx,t)=-\frac{4}{9a^2 H^2}e^{-2\zeta(\vx)}\left[\nabla^2\zeta(\vx)+\frac{1}{2}\zeta_i(\vx)\zeta^i(\vx)\right] = \frac{4}{9a^2 H^2}e^{-2\nu \sigma_0}\left[x\sigma_2-\frac{1}{2}\vec{\eta} \cdot \vec{\eta}\right].
\ee 
Now we perform the change of variables:
\be
x_{\delta} = x, \quad  \vec{\eta}_{\delta} = \vec{\eta} , \quad \nu = \frac{1}{2\sigma_0}{\rm ln}\llp\frac{4\lp x_{\delta}\sigma_2 - \frac{1}{2}\vec{\eta}_{\delta} \cdot \vec{\eta}_{\delta} \rp}{9a^2H^2\delta} \rrp.
\ee
 The argument of the logarithm is positive for $x_\delta > \vec{\eta}_{\delta} \cdot \vec{\eta}_{\delta} / 2 \sigma_2$.
The Jacobian of the transformation is given by
\be
J= \left|
\frac{1}{2\delta \sigma_0}
\right|.
\ee
Therefore the distribution in terms of the new variables is given by
\be
P(\delta, \vec{\eta}_{\delta} ,x_{\delta}) \d \delta \d^3 \eta_{\delta} \d x_{\delta} = D e^{-Q_3} 
\Theta (x_\delta \sigma_2-\eta^2_{\delta}/2)
 \d \delta \d^3 \eta_{\delta} \d x_{\delta} 
\ee
where we have defined
\be
D(\delta)= C J = \frac{6 \sqrt{3}}{8\pi^{2}\sqrt{2\pi(1-\gamma^2)}\sigma_1^3} 
\left|
\frac{1}{2\delta \sigma_0}
\right|
\ee
and
\be
2Q_3 =  \frac{1}{4\sigma_0^2}{\rm ln^2}\llp\frac{4\lp x_{\delta}\sigma_2 - \frac{1}{2}\vec{\eta}_{\delta} \cdot \vec{\eta}_{\delta} \rp}{9a^2H^2\delta} \rrp +\frac{1}{(1-\gamma^2)} \left\{ x_{\delta}- \frac{\gamma}{2\sigma_0}{\rm ln}\llp\frac{4\lp x_{\delta}\sigma_2 - \frac{1}{2}\vec{\eta}_{\delta} \cdot \vec{\eta}_{\delta} \rp}{9a^2H^2\delta} \rrp \right\}^2 +  \frac{3 \vec{\eta}_{\delta} \cdot \vec{\eta}_{\delta}}{\sigma_1^2}.
\ee
Finally, since the probability distribution is only a function of the modulus $\vec \eta_\delta \cdot \vec \eta_\delta = \eta_\delta^2$, one can change variable as $\d ^3 \eta_\delta =  \eta _\delta^2  \sin \theta_\delta \d \eta_\delta\d \theta_\delta \d \phi _\delta  $ and perform the integration on the angles which trivially results in 
\be\label{P-th}
P(\delta, \eta_\delta,x_\delta) \d \delta \d \eta_{\delta} \d x_{\delta}
 = 4 \pi \eta_\delta^2 D e^{-Q_3}  
\Theta (x_\delta \sigma_2-\eta^2_{\delta}/2)
 \d \delta \d \eta_{\delta} \d x_{\delta}.
\ee
Finally we get
\begin{tcolorbox}[colframe=white,arc=0pt]
\vspace{-.15cm}
\begin{eqnarray}
\label{kk}
\beta_\NG^\TH&=&4\pi \int_{\delta_{\rm c}} \d \delta\int_0^\infty \d\eta_\delta\,\eta_\delta^2
\int_{
\eta^2_\delta /2 \sigma_2
}^{\infty}
\d x_{\delta}\, D(\delta,x_\delta,{\eta}_\delta) e^{-Q_3}.  
\end{eqnarray}
\end{tcolorbox}
\noindent
This is an exact result, no approximations have been made at this stage\footnote{We checked that using Eq.~\eqref{kk} gives the same numerical result obtained  by  computing the probability of the overdensity integrating Eq.~\eqref{probnoch} with the insertion of a Heaviside function of the form $\Theta \lp \delta - \delta_{\rm c} \rp$ leading to the limit of integration in the variable $x$ given by the condition  $ x >(9a^2H^2/4\sigma_2){\rm exp}(2\nu\sigma_0)\delta_{\rm c}$.}.

\subsection{Spiky power spectrum}
\begin{figure}[t]
\includegraphics[width = 0.6\linewidth]{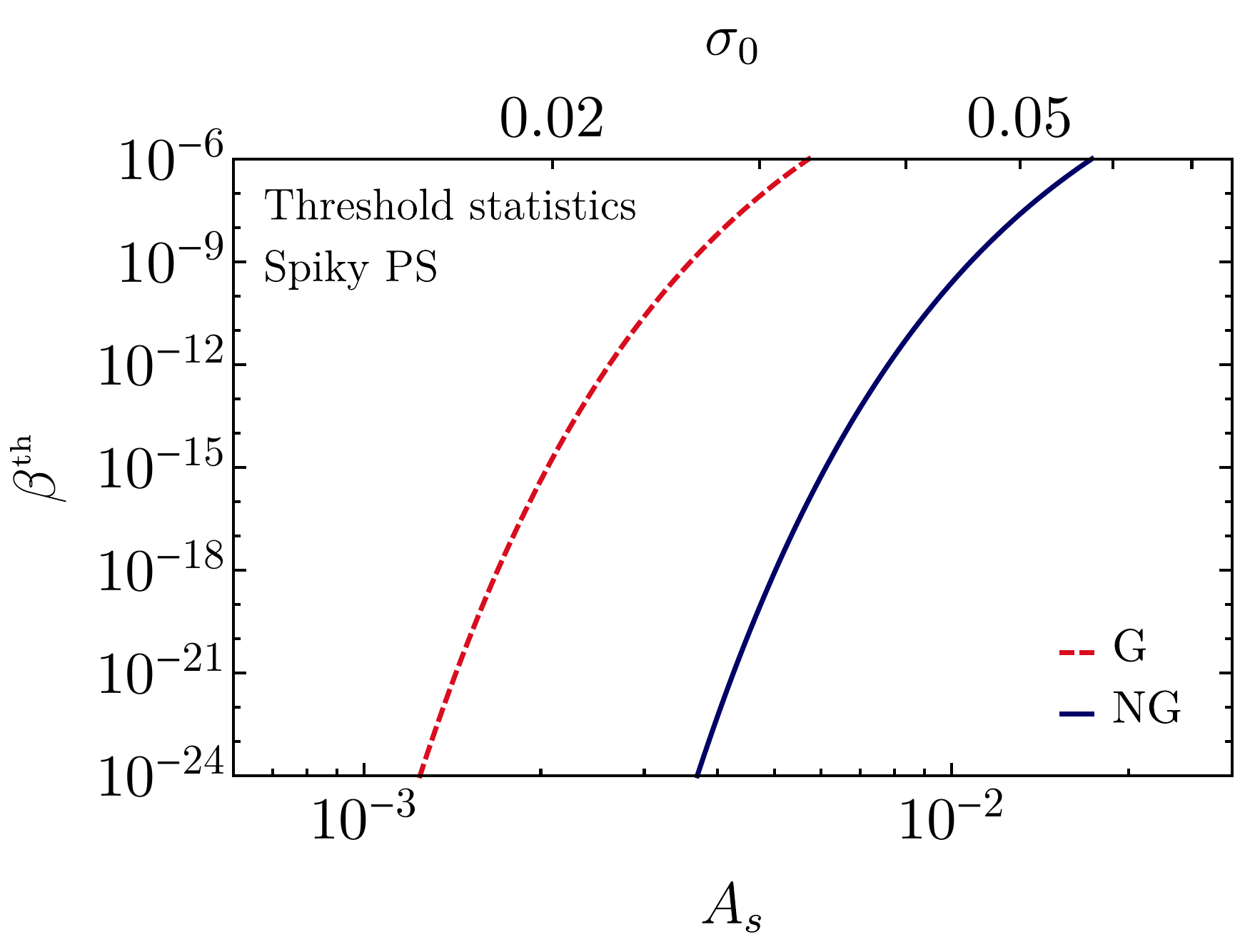}
\centering
\caption{ \it
Mass fraction $\beta^{\text{{\tiny {\rm th}}}}$ in  the case of a spiky power spectrum as a function of $A_s$, for the non-Gaussian and the gaussian cases computed using the threshold statistics. } 
\label{fig-th}
\end{figure}

In the limit of $\gamma\simeq 1$, i.e. for power spectra whose width is very narrow (typical of the PBHs), we can simplify our expressions dramatically. First of all, from Eq. 
(\ref{Q2}) one sees that the distribution in $x_\delta$ becomes a Dirac delta centered in $x_*\simeq \nu$. 
We then obtain
\be
P(\delta, \eta_{\delta} ,x_{\delta}) \d \delta \d \eta_{\delta} \d x_{\delta} 
=
 4 \pi \eta^2_\delta E e^{-Q_4} \delta_D\lp x_{\delta}- \frac{1}{2\sigma_0}{\rm ln}\llp\frac{4\lp x_{\delta}\sigma_2 - \frac{1}{2}\eta^2_{\delta} \rp}{9a^2H^2\delta} \rrp\rp 
  \Theta (x_\delta \sigma_2-\eta^2_{\delta}/2)
 {\rm d}\delta \d \eta_{\delta} {\rm d}x_\delta
\ee
where 
\be
E =  \frac{6\sqrt{3}}{8\pi^{2}\sigma_1^3} 
\frac{1}{2\delta \sigma_0 }
\qquad
\text{and}
\qquad
2Q_4 =  \frac{1}{4\sigma_0^2}{\rm ln^2}\llp\frac{4\lp x_{\delta}\sigma_2 - \frac{1}{2}\eta^2_{\delta}\rp}{9a^2H^2\delta} \rrp +  \frac{3 \eta^2_{\delta}}{\sigma_1^2}.
\ee
Then, to perform the integral in $\d \eta_{\delta}$, we rewrite the Dirac delta as 
\be
\delta_D\lp x_{\delta}- \frac{1}{2\sigma_0}{\rm ln}\llp\frac{4\lp x_{\delta}\sigma_2 - \frac{1}{2}\eta_\delta^2 \rp}{9a^2H^2\delta} \rrp\rp  = \delta_D \lp \eta_\delta - \eta^{\rm c}_\delta \rp \cdot \left | \frac{9 a^2 H^2 \delta \sigma_0}{e^{-2 \sigma_0 x_{\delta}} \sqrt{8 \sigma_2 x_{\delta}-18 a^2 H^2 \delta e^{2 \sigma_0 x_{\delta}}}} \right |,
\ee
where
\be
 \eta^{\rm c}_\delta = \sqrt{2 \sigma_2 x_{\delta}-\frac{9}{2} a^2 H^2 \delta  e^{2 \sigma_0 x_\delta}}
\ee
and where we have chosen the positive root since $\eta_\delta$ is always positive. The root imposes the condition
\be
2 \sigma_2 x_{\delta}-\frac{9}{2} a^2 H^2 \delta  e^{2 \sigma_0 x_\delta}>0,
\ee
which is solved by ($W_0$ and $W_{-1}$ are the so-called principal and negative branches of the Lambert function)
\be
x_{-}(\delta)=-\frac{1}{2\sigma_0}W_0\left(-\frac{9a^2H^2\sigma_0\delta}{2\sigma_2}\right)<x_\delta< -\frac{1}{2\sigma_0}W_{-1}\left(-\frac{9a^2H^2\sigma_0\delta}{2\sigma_2}\right)=x_{+}(\delta),
\label{xminus-xplus}
\ee
with the requirement that\footnote{The condition \eqref{requirement-delta} leaves a really narrow window in terms of $\delta$, $0<\delta<0.59$. This means that  the Dirac delta  power spectrum would not be a good choice where the threshold is larger than 0.59. Of course, such a monochromatic power spectrum is only an approximation for more physical narrow power spectra. }
\be
0<\delta<\frac{1}{e}\cdot \frac{2\sigma_2}{9a^2H^2\sigma_0} = \delta_+ \,.
\label{requirement-delta} 
\ee
After integrating in $\d \eta_{\delta}$, we find that the joint probability is
\begin{eqnarray}
P(\delta, x_{\delta}) \d \delta  \d x_{\delta} 
&=&
 \frac{54 \sqrt{3} }{8\pi\sqrt{2}} \frac{a^3 H^3}{\sigma_1^3} \sqrt{4 x_{\delta} \frac{\sigma_2}{a^2 H^2} -9  \delta e^{2 \sigma_0 x_{\delta}}} \nonumber\\
& \times & 
 \exp \llp -\frac{1}{2}x_\delta^2+ 2 \sigma_0 x_{\delta} - \frac{3a^2H^2}{4\sigma_1^2}\left( 4 x_{\delta} \frac{\sigma_2}{a^2 H^2} -9  \delta e^{2 \sigma_0 x_{\delta}}\right) \rrp  \d \delta  \d x_{\delta}.
\end{eqnarray}
This means that the threshold probability is
\begin{tcolorbox}[colframe=white,arc=0pt]
\vspace{-.15cm}
\begin{eqnarray}
\beta_\NG^\TH&=&\int_{\delta_{\rm c}}^{\delta_+} \d \delta \int_{x_{-}(\delta)}^{x_{+}(\delta)}\d x_{\delta}\,\frac{54\sqrt{3} }{8\pi\sqrt{2}} \frac{a^3 H^3}{\sigma_1^3} \sqrt{4 x_{\delta} \frac{\sigma_2}{a^2 H^2} -9  \delta e^{2 \sigma_0 x_{\delta}}} \nonumber\\
& \times & 
 \exp \llp -\frac{1}{2}x_\delta^2+ 2 \sigma_0 x_{\delta} - \frac{3a^2H^2}{4\sigma_1^2}\left( 4 x_{\delta} \frac{\sigma_2}{a^2 H^2} -9  \delta e^{2 \sigma_0 x_{\delta}}\right) \rrp,
\label{exact-betathNG} 
\end{eqnarray}
\end{tcolorbox}
\noindent
where the higher extremum of integration in $\delta$ is due to (\ref{requirement-delta}).
In Fig.~\ref{fig-th} one can find the comparison of the gaussian and non-Gaussian mass  functions computed using the threshold statistics for a spiky power spectrum.
To proceed  further and provide more analytical insights, we notice that the 
 integration over $x_\delta$ in  Eq.  (\ref{exact-betathNG}) is highly dominated by the lower extremum of integration $x_- \left( \delta \right)$. As we show in Appendix \ref{app:analytic}, the integrand in this region is very well approximated by 
\begin{eqnarray}
\beta_{\NG}^{\TH}&\simeq&\frac{54\sqrt{6} }{8\pi} \frac{a^2 H^2 \,  \sigma_2^{1/2}  }{\sigma_1^3}  \int_{\delta_{\rm c}}^{\delta_+} \d \delta \, \sqrt{ 1 - 2 \sigma_0 \, x_-\left( \delta \right) } \,  \exp \left[   \frac{x_- \left( \delta \right) \left\{ 4 \sigma_0 \sigma_1^2 + 6 \sigma_2 + x_- \left( \delta \right) \left[ \sigma_1^2-12 \sigma_0 \sigma_2  \right] \right\}}{2 \sigma_1^2} \right]  \nonumber\\
&\cdot&
\int_{x_{-} \left( \delta \right) }^{x_{+} \left( \delta \right)} \d x_{\delta}\,\sqrt{ x_\delta - x_- \left( \delta \right)   } \,  
\exp \left[ - \frac{3 \sigma_2 + x_- \left( \delta \right) \left[ \sigma_1^2 - 6  \sigma_0 \sigma_2 \right] }{\sigma_1^2} x_\delta \right].
\label{beta-integrand-approx}
\end{eqnarray}
 Since the integral in the second line is highly dominated by the lower extremum of integration, we can set $x_+(\delta) \to \infty$ and perform the integration analytically, obtaining (for $\gamma \simeq 1$) 
\begin{eqnarray}
\beta_{\NG}^{\TH}\simeq \frac{54}{8} \, \sqrt{\frac{3}{2 \pi}} \, \frac{a^2 H^2}{\sigma_2} \, \int_{\delta_{\rm c}}^{\frac{2 \sigma_2}{9 a^2 H^2 \sigma_0 \, e} }  \d \delta \, 
\frac{\sqrt{1-2 \sigma_0 \, x_- \left( \delta \right)}}{\left\{ \sigma_0 \, x_- \left( \delta \right) + 3 \left[ 1 - 2 \sigma_0 \, x_- \left( \delta \right) \right] \right\}^{3/2} } \,  {\rm e}^{-\frac{x_- \left( \delta \right) \left[ x_- \left( \delta \right) - 4 \sigma_0 \right]}{2}}.
\label{beta-result-approx} 
\end{eqnarray}
In Appendix  \ref{app:analytic}  we show that this expression is extremely accurate for the case of a Dirac delta  power spectrum of the curvature perturbation. 

We can perform the final integral (\ref{beta-result-approx}) by changing the variable of integration from $\delta$ to $x_- \left( \delta \right)$. The lower and higher extrema of integration then become, respectively, $x_-\left( \delta_{\rm c} \right)$ and $1/2 \sigma_0$. The integrand is highly dominated by the region around the lower extremum, so that we can send the higher extremum to infinity. We can also evaluate all the integrand, apart from the exponential factor ${\rm exp}(-x_-^2 \left(\delta \right)/2)$ at the lower extremum. The integral of this exponent can be then done analytically, and its result (the complementary error function)  can be expanded in the limit of large argument. This leads to 
\begin{equation}
\beta_{\NG}^{\TH} \simeq 6\sqrt{\frac{3}{2 \pi}} \, \frac{\left( 1 - 2 \sigma_0 \, x_{\rm c} \right)^{3/2}}{2 x_c \left( 3 - 5 \, \sigma_0 \, x_{\rm c} \right)^{3/2}} \, {\rm e}^{-\frac{x_{\rm c}^2}{2}} \;\;\;,\;\;\; 
x_{\rm c} \equiv x_- \left( \delta_{\rm c} \right) =   - \frac{1}{2 \sigma_0} \, W_0 \left( - \frac{9 a^2 H^2 \sigma_0 \delta_c}{2 \sigma_2} \right). 
\label{beta-result2-approx} 
\end{equation} 
The accuracy of this result is shown in Figure \ref{fig:final-res-delta} of Appendix C, performed for the case of a Dirac delta  power spectrum of the curvature perturbation, where it is compared with a two-dimensional numerical integration 
 of the starting expression (\ref{exact-betathNG}).

\noindent
\subsection{Log-normal power spectrum}
\begin{figure}[t]
\includegraphics[width = 0.6\linewidth]{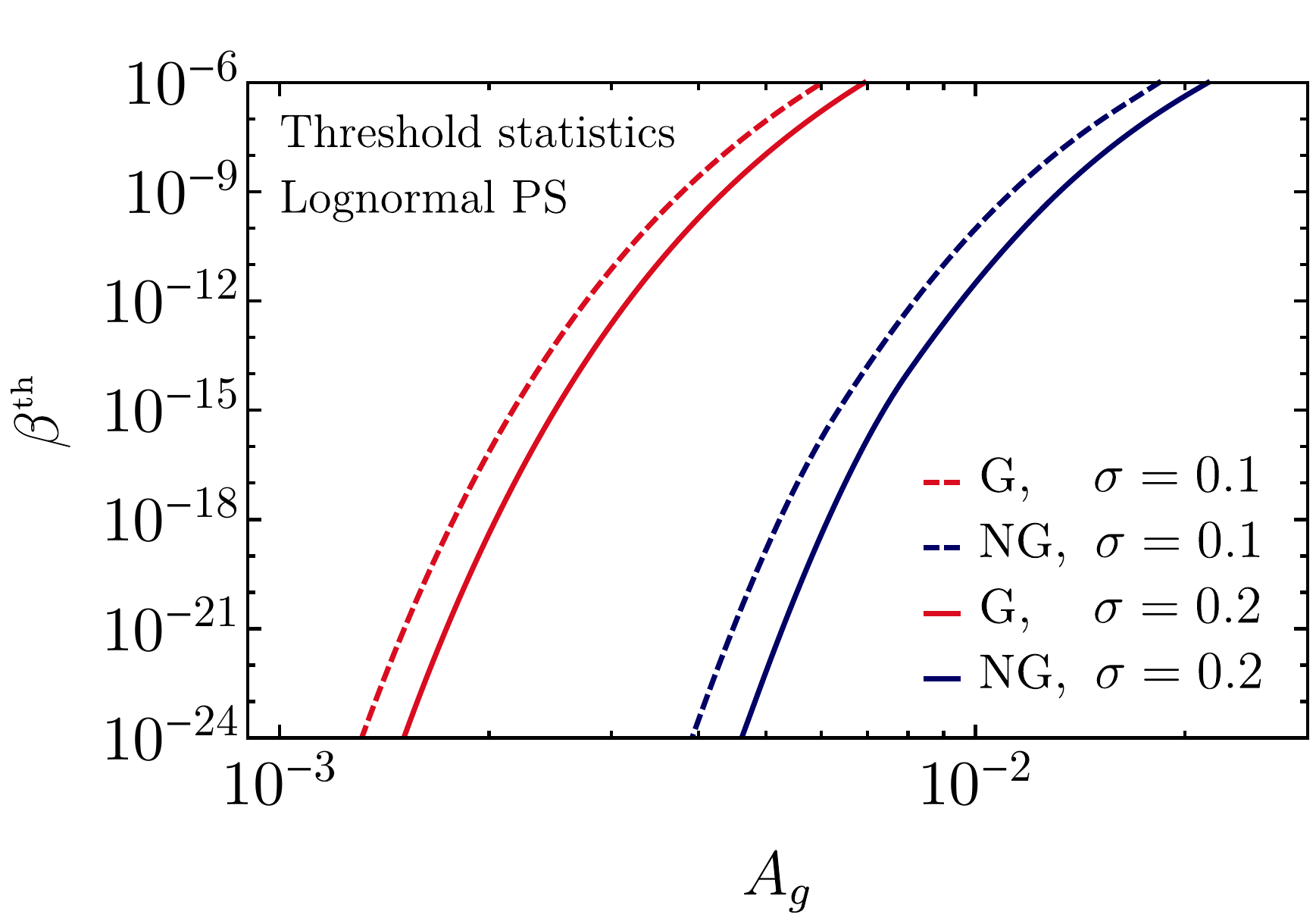}
\centering
\caption{\it
Mass fraction $\beta^{\text{{\tiny {\rm th}}}}$ as a function of $A_g$ for the log-normal power spectrum (PS)  computed using threshold statistics for both the gaussian and the non-Gaussian case.
} 
\label{figbetath}
\end{figure}

We assume again a  power spectrum of the form
\be
\mathcal{P}_\zeta(k) =\frac{A_g}{\sqrt{2 \pi } \sigma} \exp \llp -\frac{ \ln ^2\lp k/ k_\star \rp }{2 \sigma^2 }\rrp.
\ee
Then, one can integrate Eq.~\eqref{kk} numerically  to get the mass fraction.
In Fig.~\ref{figbetath} we plot the beta for various values of $\sigma$ as a function of $A_g$. 

\subsection{Broad power spectrum}
\begin{figure}[t]
\includegraphics[width = 0.61\linewidth]{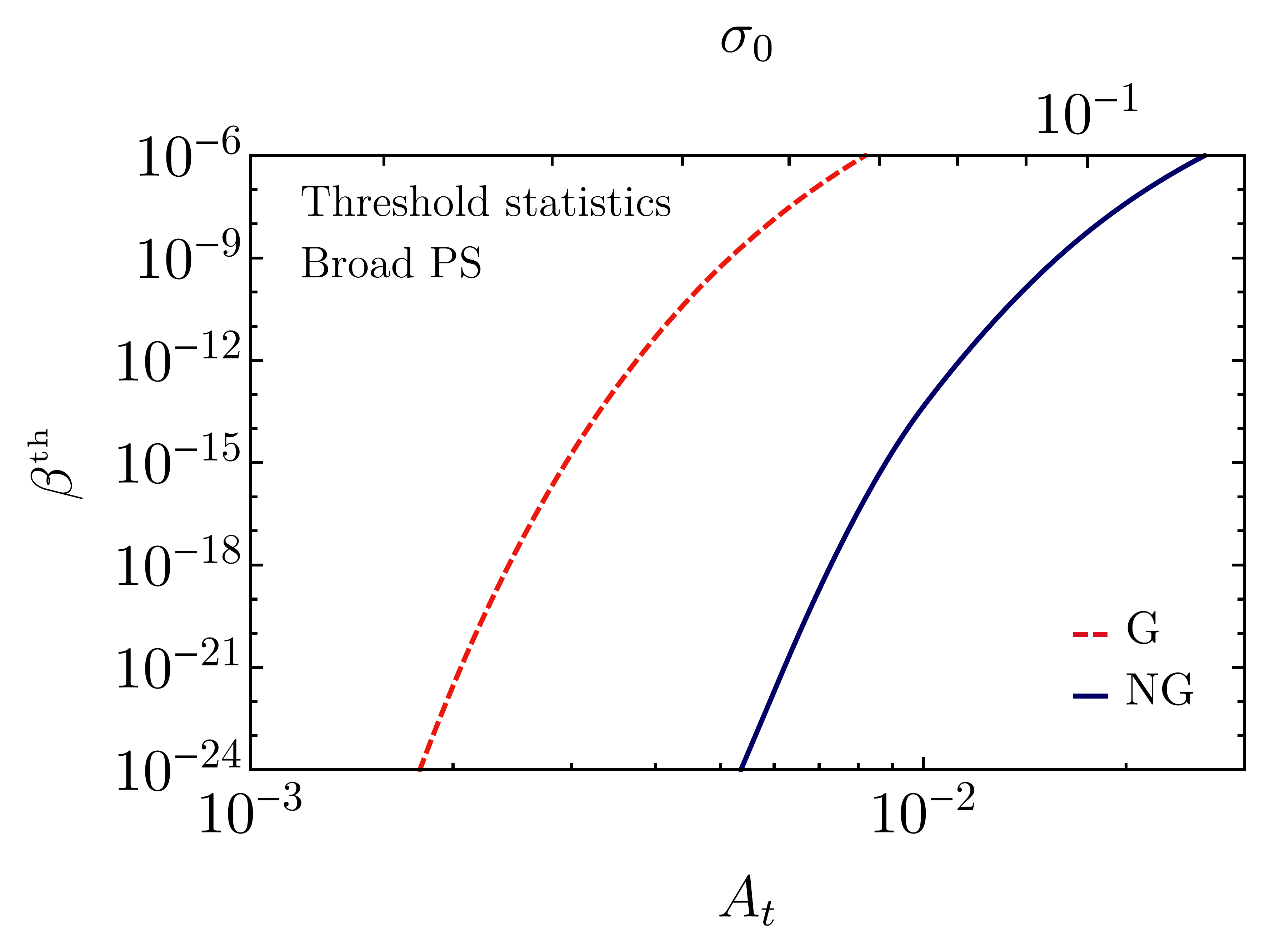}
\centering
\caption{\it
Mass fraction $\beta^{\text{{\tiny {\rm th}}}}$ as a function of $A_t$ for the broad (top-hat) power spectrum  computed using threshold statistics for both the gaussian and the non-Gaussian case.
}
 \label{fig-betath-2}
\end{figure}

We also  consider a broad  power spectrum, that is a top-hat with amplitude $A_t $ as
\be
\mathcal{P}_\zeta (k) = A_t\, \Theta (k_{\rm max}-k)\, \Theta (k-k_{\rm min}) 
\ee
where $\Theta$ stands for the Heaviside step function and $k_{\rm max}\gg k_{\rm min}$. Again, the parameters used are $k_{\rm max}\simeq 3.5/r_m$,  $\delta_{\rm c}= 0.51$ \cite{mg} and, to disregard unphysical long wavelength modes, variances are obtained by choosing $a_mH_m$ as the infrared cut-off. The results are presented in Fig. \ref{fig-betath-2}.

We conclude that threshold statistics confirms what we found in peak theory: independently from the power spectrum, non-Gaussian abundances  are smaller than the gaussian ones. 

We also  see that the difference between the gaussian and the non-Gaussian cases in terms of the amplitude of the power spectrum is about a factor $(2\div 3)$, the same for the Dirac delta case. This is the shift one should adopt if insisting in using the gaussian expressions.

\section{Conclusions}
In this paper we have discussed the impact of the non-Gaussianity arising from the non-linear relation between the density contrast and the curvature perturbation
when dealing with PBH abundances. 
We have proposed two different methods to deal with such  unavoidable and intrinsic non-Gaussianity, providing simple analytical expressions for the abundance to take it into account. 

The first method is based on peak theory and on the realisation that the number of peaks in the overdensity is approximately equal to the number of peaks in the curvature perturbation as long as one restricts her/himself
to those peaks having large spatial second derivatives at the peak location. 

The second method relies on the  threshold statistics and contains no approximations. Both methods
show that the intrinsic non-Gaussianity makes it harder to generate PBHs.  In particular, if one insists in adopting the gaussian expression for the abundance coming from threshold statistics, one has simply to increase the amplitude of the power spectrum by a factor \footnote{Our findings agree with those recently obtained in   Refs. \cite{jap,ymb}, where the PBH abundance has been derived using  the averaged (over  a volume of radius $r_m$) density fluctuation constructed out of a radial profile in the curvature perturbation $\zeta$.  Adopting the volume averaged density provides a clear relation between the linear gaussian component of the peak height and the non-linear peak height. However, in order to make use  consistently of the obtained critical threshold value, one needs to identify peaks in $\zeta$ with the peaks in $\delta$, which we have shown here is true for spiky peaks in $\zeta$. 


In Ref. \cite{haradath} the authors computed the abundance using peak theory for the comoving curvature perturbation by setting a threshold on the $\zeta$, contrarily to our choice of expressing the abundance and the threshold  in terms of the overdensity field.
}   ${\cal O}(2 \div 3 )$.

Our findings do not alleviate the differences between peak theory and threshold statistics in the computation of the abundance, already present at the gaussian level \cite{b}.

 Our results can be surely improved along some directions. It would be important to have a full non-Gaussian extension of peak theory. More importantly, the intrinsic non-Gaussianity of the overdensity is expected to  change the shape of the profile of the peaks which eventually give rise to PBHs upon collapse. Since the threshold
 $\delta_\pk^{\rm c}$ depends on the shape of the overdensity, such non-Gaussianity might change  as well the value of $\delta_\pk^{\rm c}$. We leave this study for a future  publication \cite{kmr}.

\bigskip
\bigskip
\noindent
\begin{acknowledgments}
\noindent
We thank I. Musco for  many fruitful discussions.  We thank C. Byrnes, I. Musco and S. Young for sharing their draft \cite{ymb} with us and for useful interactions.  M.P. and C.U. thank Geneva Theoretical Physics group for their hospitality during this project. V. DL. thanks the Galileo Galilei Institute for Theoretical Physics (Florence, Italy) for the nice hospitality during the realisation of this project. A.R., V. DL., and G.F. are   supported by the Swiss National Science Foundation (SNSF), project {\sl The non-Gaussian Universe and Cosmological Symmetries}, project number: 200020-178787. C.U. is supported by European Structural and Investment Funds and the Czech Ministry of Education, Youth and Sports (Project CoGraDS- CZ.02.1.01/0.0/0.0/15 003/0000437) and would like to acknowledge networking support by the COST Action GWverse CA16104.
\end{acknowledgments}

\noindent

\appendix

\renewcommand\theequation{\Alph{section}.\arabic{equation}}
\section{The cumulants for a narrow power spectrum}\label{cumulants}
\renewcommand{\theequation}{A.\arabic{equation}}

In this Appendix we derive the relations (\ref{d234c-res}) of the main text. We start from Eq. (\ref{curv}), that we need to expand as a power series of $\zeta$. We denote by  $\delta_n$ the term that is of ${\cal O } \left( \zeta^n \right)$
\begin{eqnarray} 
& & \delta_1 = - c_\star \, \partial_i \partial_i \zeta , \nonumber\\ 
& & \delta_n = c_\star \, \frac{\left( - 1 \right)^n 2^{n-1}}{\left( n - 1 \right) !} \, 
\zeta^{n-2} \left(  \zeta  \, \partial_i \partial_i \zeta - \frac{n-1}{4}  \, \partial_i \zeta \, \partial_i \zeta \right) ,
 \;\;\;\; n = 2 ,\, 3 ,\, 4 ,\, \dots \;. 
\end{eqnarray} 
Using the convention  
\begin{eqnarray}
&& \zeta \left( \vec{x} \right) = \int \frac{\d^3 p}{\left( 2 \pi \right)^3} \, {\rm e}^{i \vec{p} \cdot \vec{x} } \, \zeta \left( \vec{p} \right) , 
\end{eqnarray} 
for the Fourier transform of the curvature perturbation, and symmetrizing over the momenta $p_i$ of the Fourier modes, the above relations can be cast in the form 
\begin{align}
\delta_1 \left( 0 \right)  &= c_\star \int \frac{\d^3 p}{\left( 2 \pi \right)^3}   \, p^2 \zeta \left( \vec{p} \right) , \nonumber\\ 
\delta_n \left( 0 \right) 
&= c_\star \, 
\frac{\left( - 2\right)^{n-1}}{n!} 
\,  \prod_{k=1}^n \left[ \int \frac{\d^3 p_k}{\left( 2 \pi \right)^3 } 
\, \zeta \left( \vec p_k \right) \right] \, 
\left[  \sum_{i=1}^n p_i^2 - \frac{1}{2} \sum_{i=1}^{n-1} \sum_{j=i+1}^n  \vec{p}_i \cdot \vec{p}_j \right] ,
\;\; n = 2 ,\, 3 ,\, 4 ,\, \dots \;. 
\label{dn-expanded}
\end{align} 
We are interested in computing the connected $2$-, $3$- and $4$-point correlation functions of $\delta \left( 0 \right) = \sum_{n=1}^\infty \delta_n$, where by connected we mean terms that cannot be factorized as products of smaller-order correlation functions. Under the assumption of Gaussianity of the curvature $\zeta$, all the correlators can be broken down to the products of the two-point function of $\zeta$, 
\begin{equation}\label{dd-ps}
\left\langle \zeta \left( \vec{p} \right) \zeta \left( \vec{q} \right) \right\rangle = P_\zeta \left( p \right) \lp 2 \pi \rp ^3 \delta^{(3)} \left( \vec{p} + \vec{q} \right) = \frac{2 \pi^2}{p^3} \, {\cal P}_\zeta \left( p \right) \lp 2 \pi \rp ^3\delta^{(3)} \left( \vec{p} + \vec{q} \right) . 
\end{equation} 
The practical effect of computing a connected, rather than a full, correlator is that some of the contractions are not included. To give just one example, we have 
\begin{align}
\left\langle \delta_2^2 \left( 0 \right)  \right\rangle_c  = & 
\left\langle \delta_2^2 \left( 0 \right)  \right\rangle - 
\left\langle \delta_2 \left( 0 \right)  \right\rangle^2  \nonumber\\ 
 = & c_\star^2  
\int \frac{\d^3 p_1 \, \d^3 p_2 \, \d^3 q_1 \, \d^3 q_2}{\left( 2 \pi \right)^{12}} 
\left[ p_1^2 + p_2^2 - \frac{\vec{p}_1 \cdot \vec{p}_2}{2} \right]  
\left[ q_1^2 + q_2^2 - \frac{\vec{q}_1 \cdot \vec{q}_2}{2} \right] \left\langle 
\zeta \left( \vec{p}_1 \right)  \zeta \left( \vec{p}_2 \right)  \zeta \left( \vec{q}_1 \right)  \zeta \left( \vec{q}_2 \right) 
\right\rangle_c , 
\end{align} 
with 
\begin{equation}
\left\langle \zeta \left( \vec{p}_1 \right)  \zeta \left( \vec{p}_2 \right)  
\zeta \left( \vec{q}_1 \right)  \zeta \left( \vec{q}_2 \right) \right\rangle_c  = 
\left\langle \zeta \left( \vec{p}_1 \right)   \zeta \left( \vec{q}_1 \right)  \right\rangle \, 
\left\langle \zeta \left( \vec{p}_2 \right)   \zeta \left( \vec{q}_2 \right)  \right\rangle + 
\left\langle \zeta \left( \vec{p}_1 \right)   \zeta \left( \vec{q}_2 \right)  \right\rangle \, 
\left\langle \zeta \left( \vec{p}_2 \right)   \zeta \left( \vec{q}_1 \right)  \right\rangle , 
\end{equation} 
with the omission of the $\left\langle \zeta \left( \vec{p}_1 \right)   \zeta \left( \vec{p}_2 \right)  \right\rangle \, 
\left\langle \zeta \left( \vec{q}_1 \right)   \zeta \left( \vec{q}_2 \right)  \right\rangle $ term. 

More in general, we note that the first cumulants are related to the full correlators by 
\begin{eqnarray}
\left\langle \delta (0) \right\rangle_c &=& \left\langle \delta (0) \right\rangle ,  
\nonumber\\ 
\left\langle \delta^2(0) \right\rangle_c &=& \left\langle \delta^2 (0) \right\rangle - \left\langle \delta (0) \right\rangle^2 ,
\nonumber\\  
\left\langle \delta^3(0) \right\rangle_c &=& \left\langle \delta^3 (0) \right\rangle - 3 \left\langle \delta (0) \right\rangle   \left\langle \delta^2 (0) \right\rangle + 2 \left\langle \delta (0) \right\rangle^3 ,
 \nonumber\\  
\left\langle \delta^4(0) \right\rangle_c &=& \left\langle \delta^4 (0) \right\rangle 
- 4  \left\langle \delta (0) \right\rangle  \left\langle \delta^3 (0) \right\rangle 
- 3  \left\langle \delta^2 (0) \right\rangle^2 
+ 12  \left\langle \delta (0) \right\rangle^2  \left\langle \delta^2 (0) \right\rangle 
- 6   \left\langle \delta (0) \right\rangle^4 .
\label{eq-cumulants} 
\end{eqnarray} 
It is worth noting that only the first cumulant is affected by the average of $\delta$. In fact, the expressions 
(\ref{eq-cumulants}) show that a shift $\delta \rightarrow \delta + C$, where $C$ is a constant, only affects the first cumulant, $\left\langle \delta \right\rangle_c \rightarrow \left\langle \delta \right\rangle_c + C$, while the higher cumulants are unchanged.

Working up to cubic order in the power of $\zeta$, we compute 
\begin{eqnarray}
\left\langle \delta^2 \left( 0 \right)  \right\rangle_c &=& \left\langle \delta_1^2 \left( 0 \right)  \right\rangle_c + 
\left\langle  \delta_2^2 \left( 0 \right) + 2  \delta_1 \left( 0 \right)  \delta_3 \left( 0 \right) \right\rangle_c 
+ \left\langle  \delta_3^2 \left( 0 \right) 
+ 2  \delta_2 \left( 0 \right)  \delta_4 \left( 0 \right)  
+ 2  \delta_1 \left( 0 \right)  \delta_5 \left( 0 \right)  \right\rangle_c ,
 \nonumber\\ 
\left\langle \delta^3 \left( 0 \right)  \right\rangle_c &=&   3 \left\langle \delta_1^2   \left( 0 \right)  \delta_2 \left( 0 \right)  \right\rangle_c 
+  \left\langle  3  \delta_1^2   \left( 0 \right)  \delta_4 \left( 0 \right) 
+   6  \delta_1   \left( 0 \right)  \delta_2 \left( 0 \right)   \delta_3 \left( 0 \right) 
+   \delta_2^3  \left( 0 \right)  \right\rangle_c , 
\nonumber\\ 
\left\langle \delta^4 \left( 0 \right)  \right\rangle_c   &=&  \left\langle 6 \delta_1^2 \left( 0 \right)  \delta_2^2 \left( 0 \right) + 4 \delta_1^3 \left( 0 \right) \delta_3(0)  \right\rangle_c  , 
\label{d2c-d3c-d4c}
\end{eqnarray} 
where we have kept together terms that are of the same order in $P_\zeta$. We note that the last expression does not contain the contraction of $ \delta_1^4 \left( 0 \right)  $ as it has no connected component. 

The evaluations of the correlators in (\ref{d2c-d3c-d4c}) is tedious, but straightforward. We expand the various terms according to (\ref{dn-expanded}) and we then split the correlators in sums of connected products of 
$\left\langle \zeta \left( \vec{p} \right) \zeta \left( \vec{q} \right) \right\rangle $. Half of the integrals over momenta are then removed with the Dirac delta functions arising from Eq.~\eqref{dd-ps}.
We divide the remaining half into integrals over the magnitude of the momenta and the angles. We encounter the following nontrivial angular integrals 
\begin{eqnarray}
& & \int \d \Omega_{{\hat p}_1}  \d \Omega_{{\hat p}_2}  \d \Omega_{{\hat p}_3} \left( {\hat p}_1 \cdot {\hat p}_2 \right)^2  = \frac{64 \pi^3}{3} ,
  \nonumber\\ 
& & \int \d \Omega_{{\hat p}_1}  \d \Omega_{{\hat p}_2}  \d \Omega_{{\hat p}_3} \left( {\hat p}_1 \cdot {\hat p}_2 \right)  \left( {\hat p}_1 \cdot {\hat p}_3 \right)  = 0 ,
 \nonumber\\ 
& & \int \d \Omega_{{\hat p}_1}  \d \Omega_{{\hat p}_2}  \d \Omega_{{\hat p}_3}  
\left( {\hat p}_1 \cdot {\hat p}_2 \right)  
\left( {\hat p}_1 \cdot {\hat p}_3 \right) 
 \left( {\hat p}_2 \cdot {\hat p}_3 \right)   =  \frac{64 \pi^3}{9}  . 
\end{eqnarray} 
The explicit evaluations then give 
\begin{eqnarray}
\left\langle \delta^2 \left( 0 \right)  \right\rangle_c &=& 
 c_\star^2  \; \int \d p \, p^3 \, {\cal P}_\zeta \left( p \right) 
 \nonumber\\ 
&+&  c_\star^2  \;   \int \frac{\d p_1}{p_1}  \frac{\d p_2}{p_2}  \, {\cal P}_\zeta \left( p_1 \right) \, {\cal P}_\zeta \left( p_2 \right)    \left[ 4  p_1^4 + 4 p_2^4 + \frac{85}{6} \, p_1^2 p_2^2    \right]  \nonumber\\ 
&+&  c_\star^2 \int \frac{ \d p_1 \, \d p_2 \, \d p_3 }{p_1 p_2 p_3}  \, {\cal P}_\zeta \left( p_1 \right)  \, {\cal P}_\zeta \left( p_2 \right)  \, {\cal P}_\zeta \left( p_3 \right) \, \left[ \frac{32}{3} \left(  p_1^4 + p_2^4 + p_3^4 \right) + \frac{415}{9} \left( p_1^2 p_2^2 +  p_1^2 p_3^2 +  p_2^2 p_2^2 \right) \right] ,
 \nonumber\\ 
\left\langle \delta^3 \left( 0 \right)  \right\rangle_c &=&  
- 6 \, c_\star^3 \int \frac{\d p_1}{p_1}  \frac{\d p_2}{p_2} 
\, {\cal P}_\zeta \left( p_1 \right) \, {\cal P}_\zeta \left( p_2 \right) \, p_1^2 \, p_2^2 \left[ p_1^2 + p_2^2 \right] 
\nonumber\\ 
&-& c_\star^3 \int \frac{\d p_1 \, \d p_2  \, \d p_3  }{p_1 p_2 p_3} \, 
\left[ 46 \left(  p_1^2 + p_2^2 \right) \left(  p_2^2 + p_3^2 \right)  \left(  p_1^2 + p_3^2 \right) + \frac{577}{9} \, p_1^2 p_2^2 p_3^2  \right]  \, {\cal P}_\zeta \left( p_1 \right)  \, {\cal P}_\zeta \left( p_2 \right)  \, {\cal P}_\zeta \left( p_3 \right) ,
 \nonumber\\ 
\left\langle \delta^4 \left( 0 \right)  \right\rangle_c &=&  c_\star^4 
\int \frac{ \d p_1 \, \d p_2  \,  \d p_3}{p_1 p_2 p_3} \, 
\left[ 16 \left( p_1^4 p_2^4 +  p_1^4 p_3^4 +  p_2^4 p_3^4 \right)  + 64 p_1^2 p_2^2 p_3^2 \left( p_1^2 + p_2^2 + p_3^2 \right)  \right] 
{\cal P}_\zeta \left( p_1 \right) \, {\cal P}_\zeta \left( p_2 \right) \, {\cal P}_\zeta \left( p_3 \right) . \nonumber\\ 
\end{eqnarray} 
In the case of a very narrow power spectrum of the curvature perturbation, that can be approximated by a Dirac delta function as in Eq. (\ref{Pzeta-delta}), these expressions give the results  (\ref{d234c-res})  reported in the main text.

\section{Spiky peaks in the curvature perturbation  versus  peaks in overdensity: a numerical treatment}\label{App-peaks}
\begin{figure}[t]
\includegraphics[width = 0.495\linewidth]{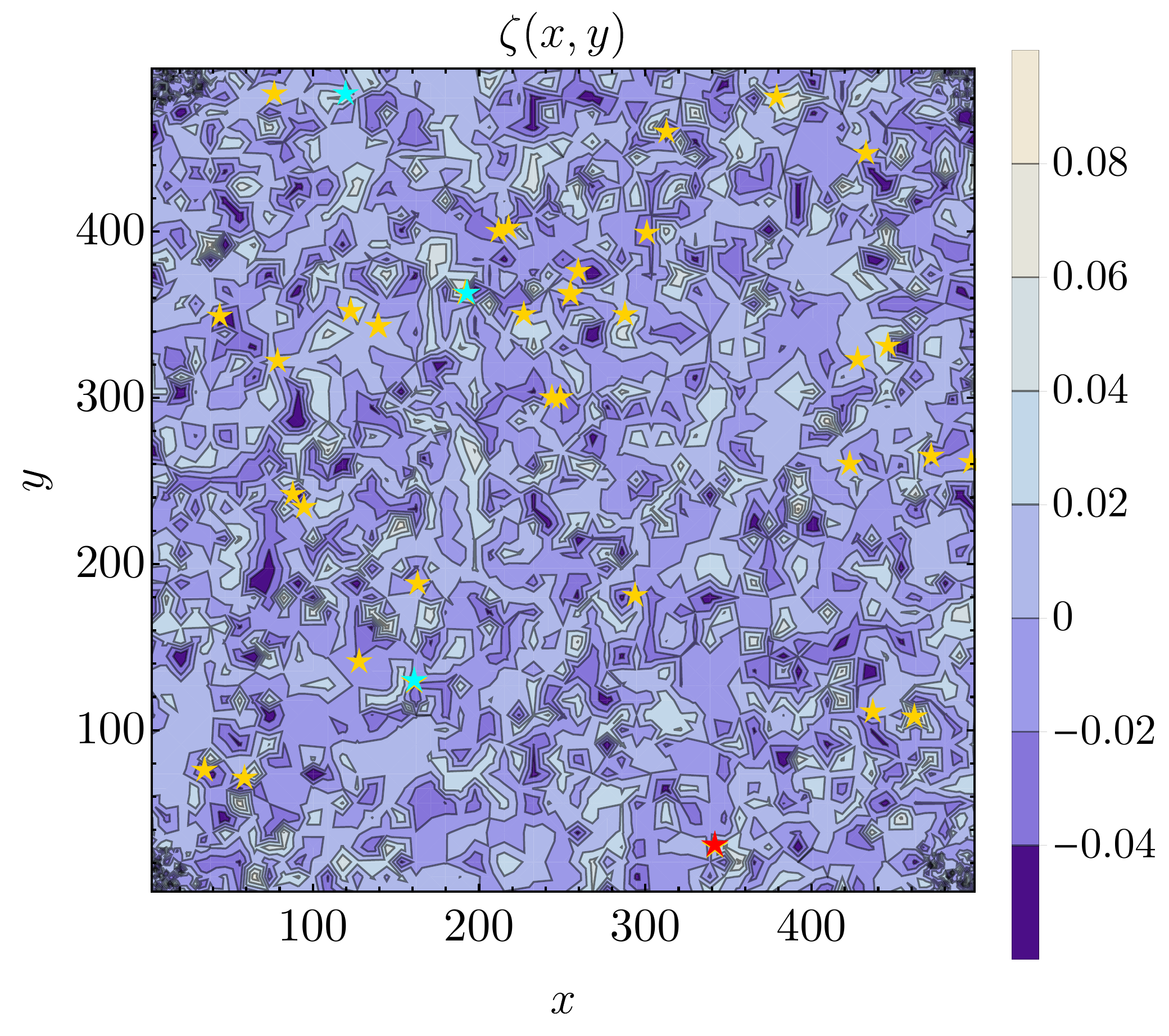}
\includegraphics[width = 0.495\linewidth]{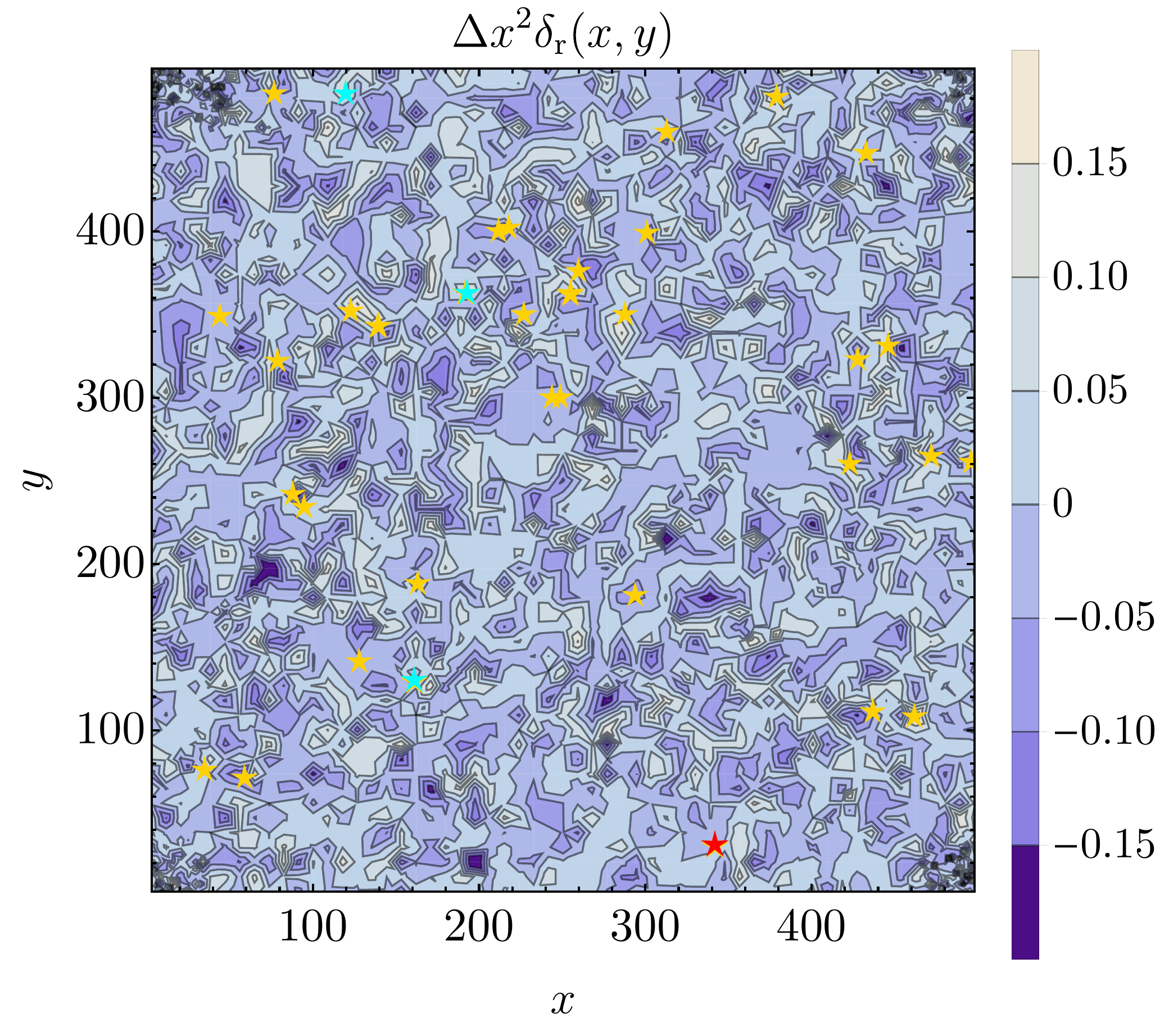}
\centering
\caption{\it
A depiction  of the two-dimensional simulation. Left: gaussian field $\zeta(x ,y)$. Right: density contrast $\delta_{\rm r}(x,y)$ found using the relation in Eq.~\eqref{dc}. The stars indicate the location of the spiky peaks in $\zeta$ and the peaks in 
$\delta_{\rm r}$, showing the location correspondence. The color code is the same as in Fig. \ref{fig32}.
} 
\label{fig2}
\end{figure}

\begin{figure}[t]
\includegraphics[width = 0.6\linewidth]{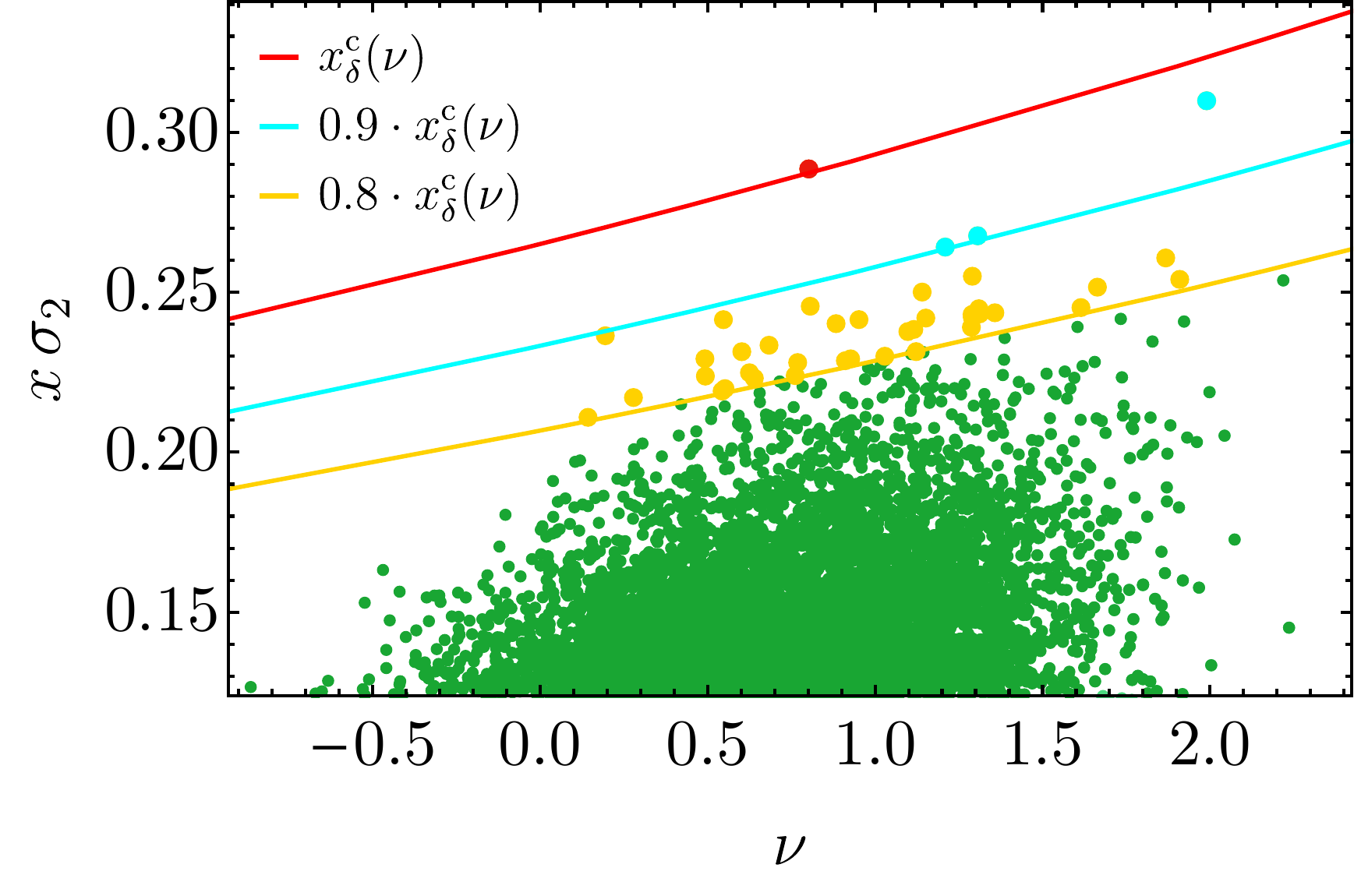}
\centering
\caption{\it
A plot  with field values of $\nu$ and $x$ (corresponding to $\zeta$ and $-\nabla^2 \zeta$) in a position of the grid. See the text for a more detailed explanation of the color code. All points are peaks in $\zeta$, but only those spiky enough are also peaks of $\delta$, as predicted.
 } 
\label{fig32}
\end{figure}

\renewcommand{\theequation}{B.\arabic{equation}}
We start from the relation between $\delta$ and $\zeta$
\be\label{dc}
\delta (\vx , t) = - \frac{4}{9} \frac{1}{a^2 H^2} e^{-2 \zeta(\vx)} \lp \nabla^2 \zeta(\vx) +\frac{1}{2} \partial_i \zeta(\vx) \partial^i \zeta(\vx) \rp 
\equiv
\frac{4}{9} \frac{1}{a^2 H^2} \delta_{\rm r}(\vx,t).
\ee
One can simulate numerically a realisation of the gaussian random field $\zeta (\vec x)$ in a $n$-dimensional  box of dimensions $N$
which is discretised using a grid of $N^n$ points with a spacing $\Delta x=1$ between them in all directions. We choose to present the analysis in a 2-dimensional  space $(n=2)$ since the results can be more easily depicted.
We set the parameters of the perturbation assuming a narrow power spectrum described by a log-normal function as
\be
\mathcal{P}_\zeta (k) = 0.01 \exp\left [ - \frac{\ln ^2 (k/k_\star)}{2 \cdot 0.1^2}\right].
\ee
The variance of the field turns out to be $\sigma_0^2=2.5 \cdot 10^{-3}$. The characteristic momentum has been chosen to be $k_\star = 0.2/\Delta x $. The realisation of the field $\zeta(\vec x)$ and the corresponding field $\delta_{\rm r}(\vec x)$ can be seen in Fig.~\ref{fig2}. There the stars indicate the location of the spiky peaks in $\zeta$ and the peaks in 
$\delta_{\rm r}$, showing the location correspondence. The color code is the same as in Fig. \ref{fig32}.

In Fig.~\ref{fig32} one can find an analysis of the field values obtained in the simulation. 
More in detail,  each point of the plot represents a peak in $\zeta$ with the corresponding values of  the rescaled amplitude $\nu$ and the curvature $x$.    The red, cyan and yellow  lines correspond to lower bounds on  $x>x_\delta^{\rm c} (\nu)$ in terms of the absolute maximum of the density contrast $\delta_{\max}$ in the simulation as
\be
x_\delta^{\rm c} (\nu) =    \frac{9a^2H^2}{4\sigma_2}e^{ 2\sigma_0\nu}  \delta_{\rm max},
\ee
with $\delta_{\rm max}=0.4$. This bound corresponds to the condition \eqref{cond}. With red, cyan and yellow dots we highlight the points which, at the same positions, have a peak in $\delta$, with $x$ satisfying the corresponding lower limits. Green dots are peaks in $\zeta$ as well, but they do not satisfy these conditions. This shows the correspondence between peaks of $\zeta$ and peaks of $\delta$, provided the condition \eqref{cond} is met. We expect that this correspondence will be even more satisfied when rarer events are simulated. We also checked that, by extending the simulation to three dimensions, and these findings are confirmed.

These results strongly indicates that,  assuming condition \eqref{cond}, peaks in $\delta$ are located at the positions of peaks in $\zeta$. 

\section{Analytic integration of the PBH abundance for spiky power spectra using threshold statistics}\label{app:analytic} 
\renewcommand{\theequation}{C.\arabic{equation}}
In this appendix we derive the expressions (\ref{beta-integrand-approx}) and (\ref{beta-result-approx}) of the main text. We start from Eq. (\ref{exact-betathNG}). One can verify that the  integration over $ x_\delta$ of this equation is highly dominated by the lower extremum $x_- \left( \delta \right)$ (from now on, in this appendix, we do not write the dependence of $x_-$ on $\delta$ to shorten the notation). We therefore perform an expansion of the integrand for $x_\delta \simeq x_- $  that allows us to perform the integration analytically. We expand the expression in the square root and in the exponent by linearising the exponential in $x_\delta - x_-$
\begin{eqnarray} 
4 x_{\delta} \frac{\sigma_2}{a^2 H^2} -9  \delta e^{2 \sigma_0 x_{\delta}} &\simeq&  4 x_{\delta} \frac{\sigma_2}{a^2 H^2} -9  \delta e^{2 \sigma_0 x_-} \left[ 1+ 2 \sigma_0 \left( x_\delta - x_- \right) \right] \nonumber\\ 
&=& \frac{4 \left( 1 - 2 \sigma_0 x_- \right) \, \sigma_2}{a^2 H^2} \left( x_\delta - x_- \right), 
\end{eqnarray} 
where the second line has been obtained exploiting the fact that $x_-$ satisfies (exactly) $\delta {\rm e}^{2 \sigma_0 x_-} = 4 \sigma_2 x_-/9 a^2 H^2$. We also approximate the first two terms in the exponent of Eq. (\ref{exact-betathNG}) as 
\begin{equation}
- \frac{1}{2} x_\delta^2 + 2 \sigma_0 x_\delta \simeq \frac{x_- \left( x_- + 4 \sigma_0 \right)}{2} - x_- x_\delta, 
\end{equation}
where we have linearised the first term on the left-hand side  to first order in $x_\delta - x_-$, while in the second term we simply put $x_\delta = x_-$ (since this term is highly subdominant). With these approximations, the expression (\ref{exact-betathNG}) reduces to the form (\ref{beta-integrand-approx}) written in the main text. 

The  integration  over $x_\delta$ in Eq. (\ref{beta-integrand-approx}) is highly dominated by the lower extremum of integration, and we can set $x_+ \to \infty$. In this way the integration can be done analytically, leading to 
\begin{eqnarray}
\beta_{\NG}^\TH \simeq \frac{18}{8} \, \frac{1}{\sqrt{2 \pi}} \, \frac{a^2 H^2}{\sigma_2} \, \left( \frac{3 \sigma_0 \sigma_2}{\sigma_1^2} \right)^{3/2} \, \int_{\delta_{\rm c}}^{\delta_+} \d \delta \, 
\frac{\sqrt{1-2 \sigma_0 x_-}}{\left( \sigma_0 x_- + \frac{3 \sigma_0 \sigma_2}{\sigma_1^2} \left( 1 - 2 \sigma_0 x_- \right) \right)^{3/2} } \,  {\rm e}^{-\frac{x_- \left( x_- - 4 \sigma_0 \right)}{2}} 
\end{eqnarray}
Recalling that these results are valid for $\gamma \equiv \frac{\sigma_1^2}{\sigma_0 \, \sigma_2} \simeq 1$
then leads to the expression (\ref{beta-result-approx}) written in the main text. 

\begin{figure}[tbp]
\centering 
\includegraphics[width=.6\textwidth]{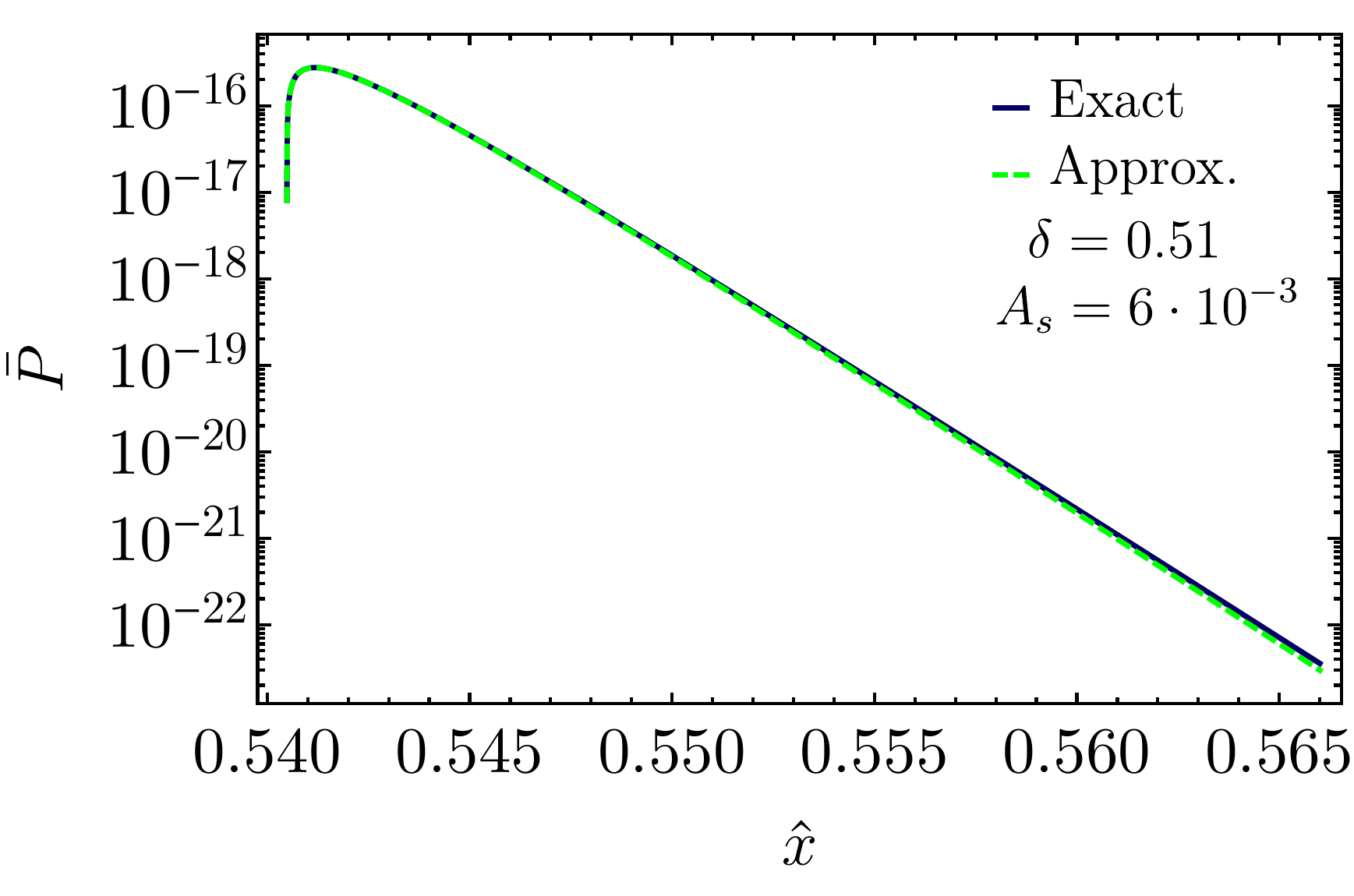}
\caption{\it 
Validity of the analytic result (\ref{res-pbty-delta}),  in the case of a Dirac delta  power spectrum of $\zeta$. 
We show the normalised probability ${\bar P} \equiv P / \left(25/9 \pi \sqrt{3} w^3A_s^{3/2} \right)$ for   $\delta = 0.51$ and $A_s = 6 \cdot 10^{-3}$. 
}
\label{fig:approx-Pbar}
\end{figure}

In the case of a Dirac delta power spectrum of the curvature perturbation $\zeta$, see Eq. (\ref{Pzeta-delta}), we have $\sigma_i = w \sqrt{ A_s} \, k_\star^i$, where $w=W(k_\star,r_m)$. Recalling that  $k_\star \simeq (27/10) a_m \, H_m$, the probability distribution reduces to 
\begin{equation} 
P \left( \delta ,\, x_\delta \right)  \simeq  
\frac{25}{9 \pi \sqrt{3} w^3 A_s^{3/2}} \, 
\sqrt{1-{\hat x}_-} \, {\rm e}^{\frac{\left( 12 + 8 w^2 A_s - 11 {\hat x}_- \right) {\hat x}_-}{8 w^2 A_s}} \, \sqrt{{\hat x} - {\hat x}_-} \, {\rm e}^{-\frac{6 - 5 {\hat x}_-}{4 w^2 A_s} \, {\hat x}}, 
\label{res-pbty-delta}
\end{equation} 
where on the right-hand side   we have defined $ {\hat x} \equiv  2  w \sqrt{A_s} x_\delta $ and 
$ {\hat x}_- \equiv  2 w  \sqrt{A_s} x_- =  - W_0 \left( - 50\delta/81\right) $ (which is the expression of the first root in eq. (\ref{xminus-xplus}) in the present case). Figure \ref{fig:approx-Pbar} confirms the validity of this result. The probability in the figure is shown for ${\hat x} \simeq {\hat x}_- \simeq 0.54$ (for the value of $\delta$ chosen in the figure), while  $ {\hat x}_+ \simeq 1.67$. We note that indeed this expression is highly dominated by the lower bound ${\hat x} \simeq {\hat x}_-$ (this extends also for the values of ${\hat x}$ not shown in the figure).

\begin{figure}[tbp]
\centering 
\includegraphics[width=.48\textwidth]{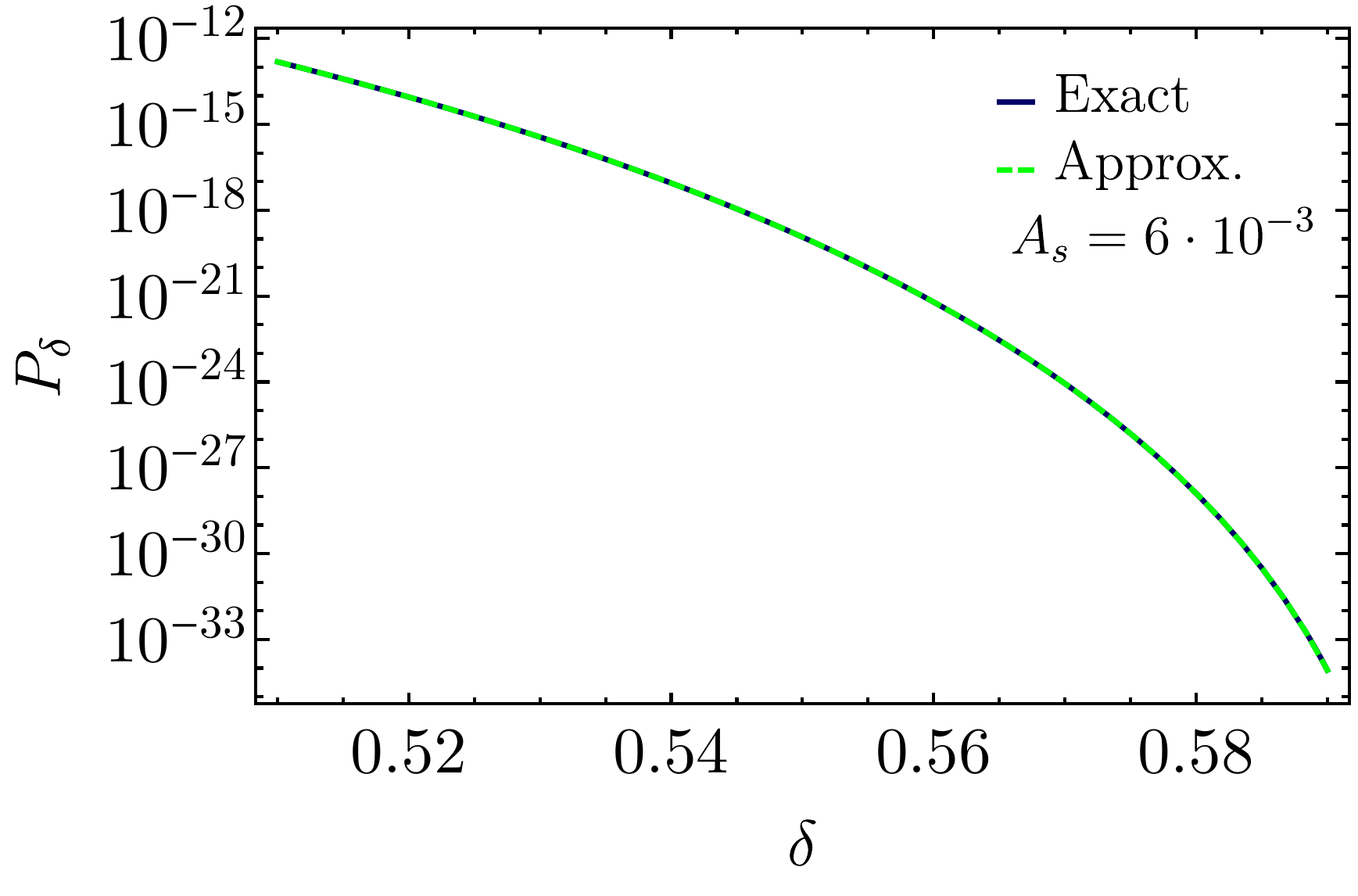}
\includegraphics[width=.5\textwidth]{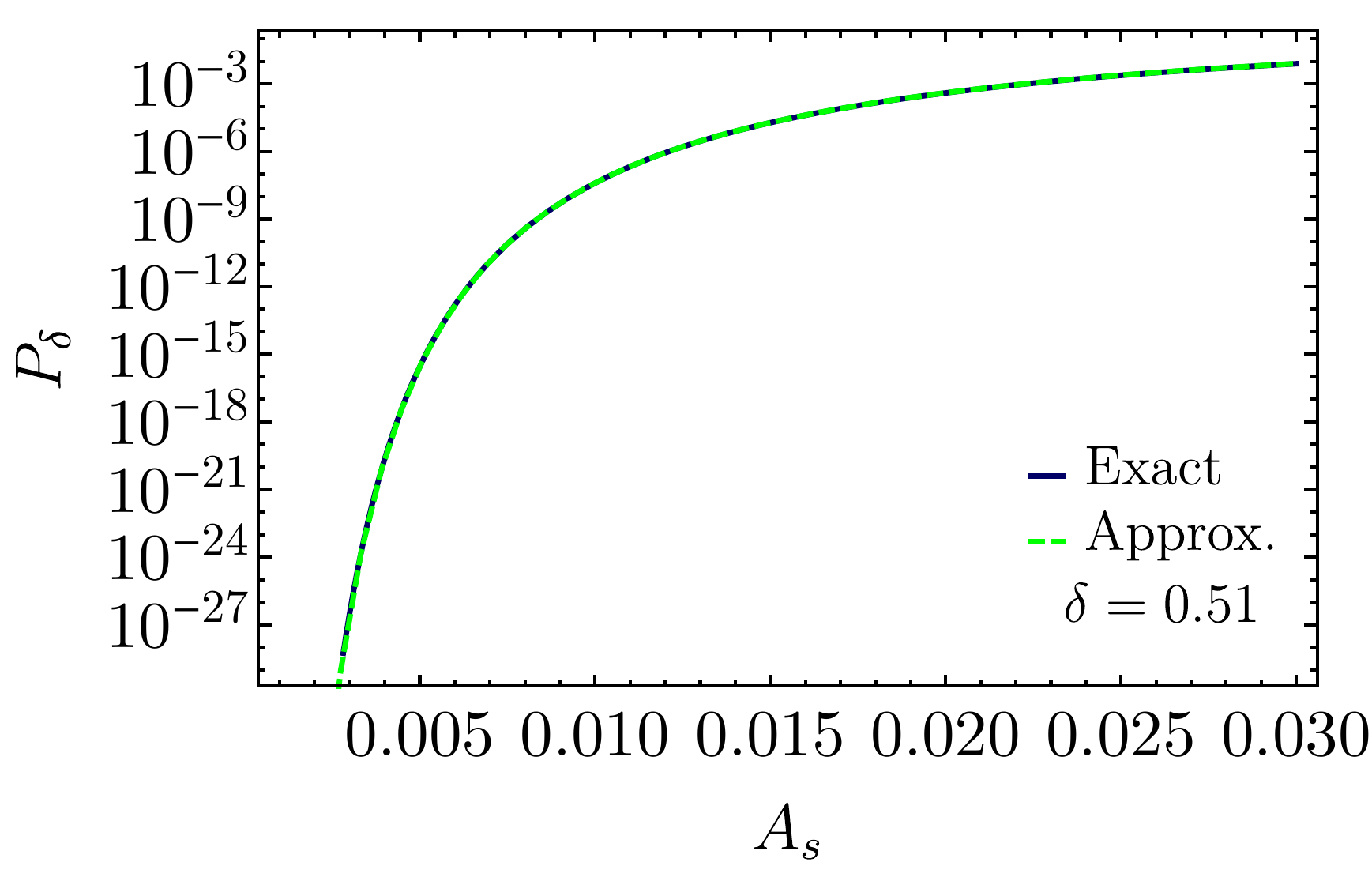}
\caption{\it 
Validity of the analytic result (\ref{res-integral-delta}) in the case of a Dirac delta  power spectrum of $\zeta$. 
This result is compared with the exact numerical integration of the expression (\ref{exact-betathNG}). 
Left: we fix $A_s = 6 \cdot 10^{-3}$ and we vary $\delta$. 
Right: we fix $\delta = 0.51$ and we vary $A_s$.  
}
\label{fig:Pdelta-As}
\end{figure}

The  integration over $x_\delta$ of this expression leads to 
\begin{equation}
\beta_{\NG}^{\TH} \simeq  
\int_{\delta_{\rm c}}^{81/50e} \d \delta \int_{x_-}^{x_+} \d x_\delta P \left(  \delta ,\, x_\delta \right)  \simeq \frac{50}{9 \sqrt{3 \pi} } \, \frac{1}{w \sqrt{A_s}} \,  \int_{\delta_{\rm c}}^{81/50e} \d \delta  \frac{\sqrt{1- {\hat x}_-}}{\left( 6 - 5 \, {\hat x}_- \right)^{3/2}} \, 
{\rm e}^{-\frac{{\hat x}_-^2}{8 w^2A_s} + {\hat x}_-}, 
\label{res-integral-delta}
\end{equation} 
where we stress that ${\hat x}_-$ depends on $\delta$. The higher extremum of integration is the upper bound in Eq.  (\ref{requirement-delta}) written in the present context. This result is extremely accurate, as we show in Figure \ref{fig:Pdelta-As}.

\begin{figure}[tbp]
\centering 
\includegraphics[width=.495\textwidth]{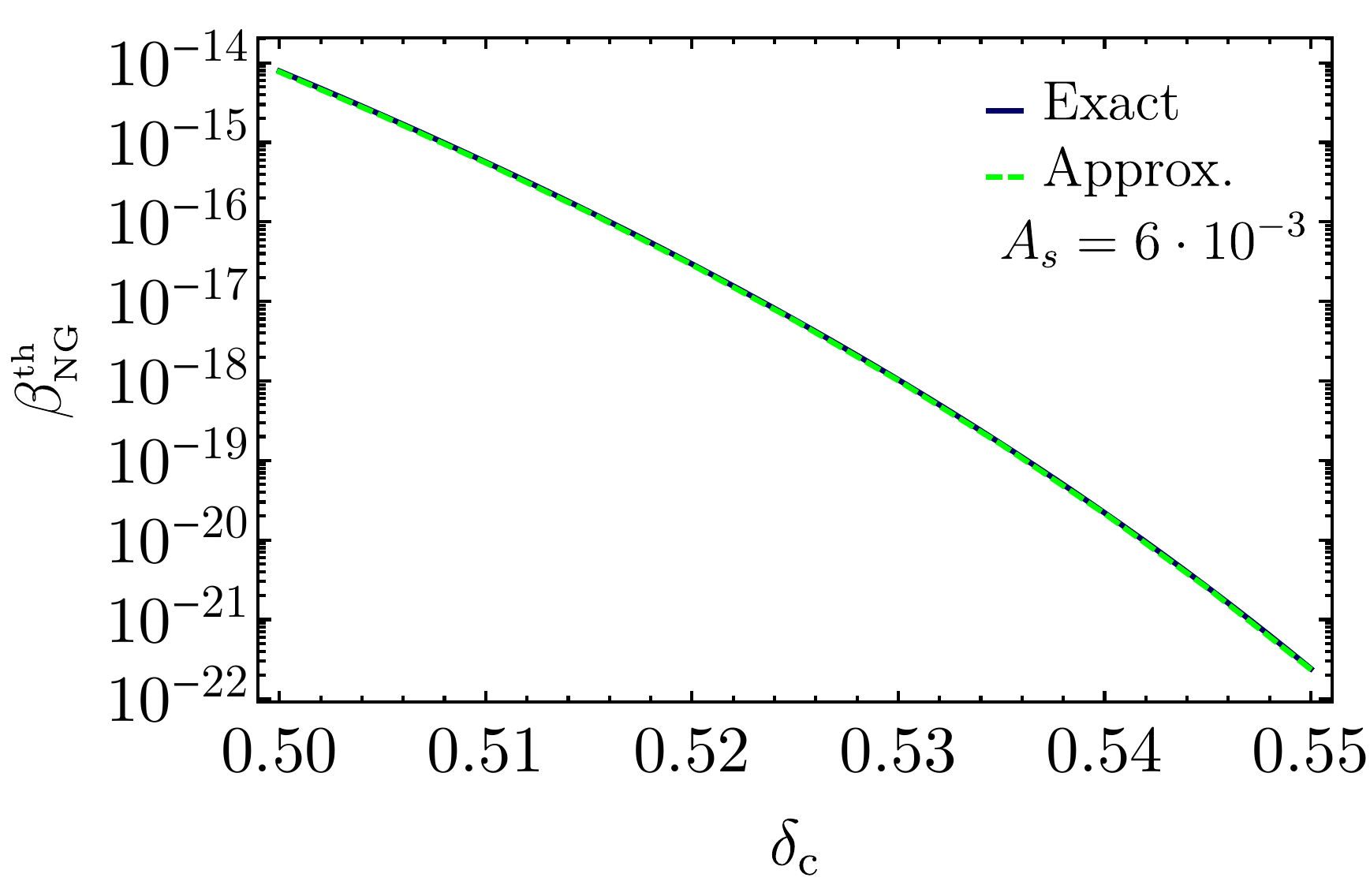}
\includegraphics[width=.485\textwidth]{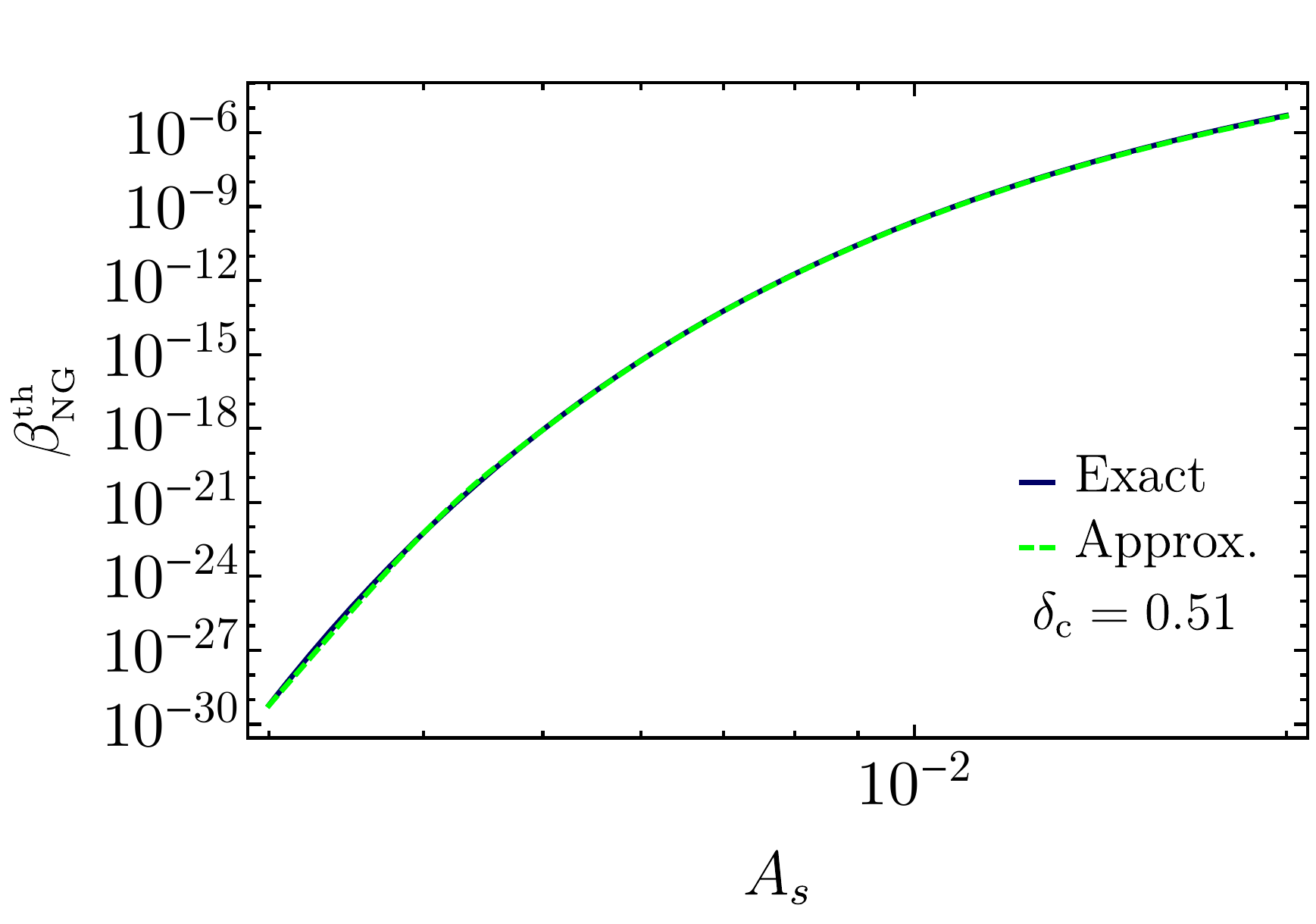}
\caption{\it 
Validity of the analytic result (\ref{res2-integral-delta}) in the case of a Dirac delta power spectrum of $\zeta$. 
This result is compared with the exact  two-dimensional  numerical integration of the expression (\ref{exact-betathNG}). 
Left panel: we fix $A_s = 6 \cdot 10^{-3}$ and we vary $\delta_{\rm c}$. Right panel: we fix $\delta_{\rm c} = 0.51$ and we vary $A_s$. }
\label{fig:final-res-delta}
\end{figure}

The expression (\ref{res-integral-delta}) can be integrated, proceeding as we did in the main text to obtain the result (\ref{beta-result2-approx}) from (\ref{beta-result-approx}). We obtain 
\begin{eqnarray} 
\beta_{\NG}^{\TH}   \simeq   12 \sqrt{\frac{3}{\pi}} \, \left( \frac{ 1 - {\hat x}_{\rm c} }{  6 - 5 {\hat x}_{\rm c} } \right)^{3/2} \, \frac{w\sqrt{A_s}}{{\hat x}_{\rm c}} \, {\rm e}^{-\frac{{\hat x}_{\rm c}^2}{8  w^2 A_s}}   \;\;,\;\; 
{\hat x}_{\rm c} \equiv {\hat x}_-\left(\delta_{\rm c} \right) =   - W_0 \left( - \frac{50 \, \delta_c}{81} \right). 
\label{res2-integral-delta}
\end{eqnarray} 
This expression also follows immediately from (\ref{beta-result2-approx}), in the limit of Dirac delta power spectrum of the curvature perturbation, and noting that $ x_{\rm c} = {\hat x}_{\rm c}/2 \, \sigma_0
 = {\hat x}_{\rm c}/2 \, w \sqrt{A_s}$. The high accuracy of this result is shown in Figure \ref{fig:final-res-delta}, where we compare it with a fully numerical two-dimensional  integration of the starting expression (\ref{exact-betathNG}).

\section{Peaks versus thresholds}\label{App-peaks}
\renewcommand{\theequation}{D.\arabic{equation}}
\noindent
In the past literature PBHs have been identified either with peaks  or with thresholds of the superhorizon overdensity, where by thresholds one means those regions in real space where the value of the density contrast is larger than a given threshold, in our case the critical value $\delta_{\rm c}$. 
Regions  characterised by large thresholds  of the overdensity are 
indeed probable to be also local extrema. We first  find 
 the average threshold statistics profile $\overline{\delta}(r)$ of the density contrast $\delta(r)$
at a given distance $r$ from the point $r=0$  (therefore without threshold)  in the following way
\be
\overline{\delta}(r)=\langle \delta(r)|\delta_0>\nu\sigma_\delta\rangle=\int_{-\infty}^{\infty}{\rm d}\delta(r) \,\delta(r) P(\delta(r)|\delta_0>\nu\sigma_\delta), 
\ee
where
\begin{eqnarray}
P(\delta(r)|\delta_0>\nu\sigma_\delta)&=&\frac{P(\delta(r),\delta_0>\nu\sigma_\delta)}{P(\delta_0>\nu\sigma_\delta)}
\end{eqnarray}
and $\delta_0 = \delta(0)$.
If both $\delta(r)$ and $\delta_0$ are Gaussian variables, one can calculate the above quantity by recalling that 
$P(\delta(r),\delta_0)$ is constructed in the standard way through the covariance matrix
\begin{eqnarray}
P(\delta(r),\delta_0)&=&\frac{1}{2\pi\sqrt{\det C}}\exp\left(-\vec{\delta}^T C^{-1}\vec\delta/2\right)\,\nonumber\\
\vec{\delta}^T&=&(\delta_0,\delta(r)),\nonumber\\
C&=&\left(\begin{array}{cc}
\sigma_\delta^2&\xi_2(r)\\
\xi_2(r) & \sigma_\delta^2\end{array}\right),
\end{eqnarray}
where
\be
\xi_2(r)=\langle\delta(\vec r)\delta(\vec 0)\rangle
\ee
is the two-point correlator in coordinate space.
From these expressions we derive
\begin{eqnarray}
P(\delta(r),\delta_0>\nu\sigma_\delta)&=&\frac{e^{-\delta^2(r)/2\sigma_\delta^2}}{2\sqrt{2\pi}\sigma_\delta}\left(1+{\rm Erf}\left[\frac{\left(\xi_2(r)\delta(r)-\nu\sigma_\delta^3\right)}{\sigma_\delta\sqrt{2\det C}}\right]\right),\nonumber\\
P(\delta_0>\nu\sigma_\delta)&=&\frac{1}{2}{\rm Erfc}\left(\nu/\sqrt{2}\right),
\end{eqnarray}
where ${\rm Erfc}(x)$ is the complementary error function.
Combining the different terms we finally get
\be
\label{fg}
\overline{\delta}(r)=\frac{\xi_2(r)}{\sigma_\delta}\sqrt{\frac{2}{\pi}}\frac{e^{-\nu^2/2}}{{\rm Erfc}\left(\nu/\sqrt{2}\right)}.
\ee
Using the expansion for large values of the argument 
\be
{\rm Erfc}\left(x\gg 1\right)\approx \frac{e^{-x^2}}{x\sqrt{\pi}},
\ee
we can finally evaluate the average $\overline{\delta}(r)$ at distance $r$ from the threshold for $\nu\gg 1$
\be
\overline{\delta}(r)\simeq \nu \,\frac{\xi_2(r)}{\sigma_\delta}.
\ee
Taking $\nu=\delta_\pk/\sigma_\delta$ one finds \
\be
\overline{\delta}(r)=\delta_\pk\frac{\xi_2(r)}{\sigma_\delta^2},
\ee
which is exactly the average profile derived in peak theory \cite{bbks}\footnote{For the non-Gaussian extension of this result, see Ref. \cite{kmr}.}. This already suggests    that large thresholds overdensity should correspond to extrema. 
To have  further evidence, we follow Ref. \cite{hof} and consider the curvature of the large threshold regions. 
 The  mean value of the second derivative of $\delta(r)$ in any random direction at $r = 0$ is (by expanding the density contrast around the origin in powers of $r$ and taking the mean value of it) with $\delta(0)=\delta_\pk$
\be
\Big<\left.\frac{{\rm d}^2{\delta}(r)}{{\rm d} r^2}\right|_{r=0}\Big>=\frac{\delta_\pk}{\sigma_\delta^2}\left.\frac{{\rm d}^2\xi_2(r)}{{\rm d} r^2}\right|_{r=0}.
\ee
The scatter of the second derivative from its mean value is found by averaging over all ${\rm d}^2\delta(r)/{\rm d}r^2|_{r=0}$ and $\delta_\pk$, yet keeping
$\delta(0)=\delta_\pk$,
\begin{eqnarray}
\Sigma_2^2&=&\Big<\left[\left.\frac{{\rm d}^2{\delta}(r)}{{\rm d} r^2}\right|_{r=0}-\frac{\delta_\pk}{\sigma_\delta^2}\left.\frac{{\rm d}^2\xi_2(r)}{{\rm d} r^2}\right|_{r=0}\right]^2\Big>\nonumber\\
&=&\left.\frac{{\rm d}^4\xi_2(r)}{{\rm d} r^4}\right|_{r=0}-\frac{1}{\sigma_\delta^2}\left(\left.\frac{{\rm d}^2\xi_2(r)}{{\rm d} r^2}\right|_{r=0}\right)^2.
\end{eqnarray}
We then get  
\be
\label{a}
\left|\frac{1}{\Sigma_2}\Big<\left.\frac{{\rm d}^2{\delta}(r)}{{\rm d} r^2}\right|_{r=0}\Big>\right|\sim\left(\frac{\delta_\pk}{\sigma_\delta^2}\frac{\sigma^2_\delta}{r_m^2}\right)\left(
\frac{\sigma_\delta^2}{r_m^4}\right)^{-1/2}=\frac{\delta_\pk}{\sigma_\delta}, 
\ee
where we have taken $\xi_2(0)\sim \sigma_\delta^2$ and  assumed  that the profile
varies over a characteristic scale $r_m$.  The condition to have large threshold, that is $\delta_\pk\gg\sigma_\delta$, implies that large threshold regions are most likely
to be local extrema, that is peaks. 

\end{document}